\documentclass[11pt,preprint,prd,aps,superscriptaddress,nofootinbib,a4paper]{revtex4}
\usepackage{graphicx,amsmath,amssymb}

\usepackage{color}
\definecolor{purple}{rgb}{0.5,0,0.5}
\definecolor{blue}{rgb}{0.0,0,0.9}
\usepackage[colorlinks=true, pdfstartview=FitV, citecolor= purple, linkcolor = blue,urlcolor=blue]{hyperref}

\usepackage[normalem]{ ulem }
\usepackage{soul}

\def\DKpp{D^+ \to K^- \pi^+ \pi^+}
\def\Dppp {D^+ \to \pi^- \pi^+ \pi^+}

\def\beq {\begin{equation}}
\def\eeq {\end{equation}}
\def\bea {\begin{eqnarray}}
\def\eea {\end{eqnarray}}

\newcommand{\ket}[1]{\left|#1\right\rangle}

\newcommand{\sand}[3]{\langle#1|#2|#3 \rangle}

\begin{document}

\title{Parametrizations of three-body hadronic  \mbox{\boldmath$B$}- and  \mbox{\boldmath$D$}-decay amplitudes in terms of analytic and unitary meson-meson form factors}

\author{D. Boito}
\affiliation{ Instituto de F\'isica de S\~ao Carlos, Universidade de S\~ao Paulo, CP 369, 13560-970, S\~ao
Carlos, SP, Brazil}

\author{J.-P.~Dedonder}
\affiliation{Sorbonne Universit\'es, Universit\'e Pierre et Marie Curie, Sorbonne Paris Cit\'e, Universit\'e
 Paris Diderot, et IN2P3-CNRS, UMR 7585, Laboratoire de Physique Nucl\'eaire et de Hautes \'Energies,  
4 place Jussieu, 75252 Paris, France}

\author{B. El-Bennich}
\affiliation{Laborat\'orio de F\'isica Te\`orica e Computacional, Universidade Cruzeiro do Sul, Rua Galv\~ao Bueno 868, 01506-000 S\~ao Paulo, SP, Brazil}

\author{R.~Escribano}
\affiliation{Grup de F\'isica Te\`orica (Departament de F\'isica) and Institut de F\'isica d'Altes Energies (IFAE), Universitat Aut\`onoma de Barcelona, E-08193 Bellaterra (Barcelona), Spain}

\author{R.~Kami\'nski}
\affiliation{Division of Theoretical Physics, The Henryk Niewodnicza\'nski Institute of Nuclear Physics,
                  Polish Academy of Sciences, 31-342 Krak\'ow, Poland}

\author{L.~Le\'sniak} 
\affiliation{Division of Theoretical Physics, The Henryk Niewodnicza\'nski Institute of Nuclear Physics,
                  Polish Academy of Sciences, 31-342 Krak\'ow, Poland}

\author{B.~Loiseau}
\affiliation{Sorbonne Universit\'es, Universit\'e Pierre et Marie Curie, Sorbonne Paris Cit\'e, Universit\'e
 Paris Diderot, et IN2P3-CNRS, UMR 7585, Laboratoire de Physique Nucl\'eaire et de Hautes \'Energies,  4 place Jussieu, 75252 Paris, France}

\date{\today }

\begin{abstract}

We introduce parametrizations of hadronic three-body $B$ and $D$ weak
decay amplitudes that can be readily implemented in experimental
analyses and are a sound alternative to the simplistic and widely used
sum of Breit-Wigner type amplitudes, also known as the isobar model.
These parametrizations can be particularly useful in the 
 interpretation of CP asymmetries in the Dalitz plots. They are
derived from previous calculations based on a quasi-two-body
factorization approach in which two-body hadronic final state
interactions are fully taken into account in terms of unitary $S$- and
$P$-wave $\pi\pi$, $\pi K$ and $K \bar K$ form factors. These form
factors can be determined rigorously, fulfilling fundamental
properties of quantum field-theory amplitudes such as analyticity and
unitarity, and are in agreement with the low-energy behaviour
predicted by effective theories of QCD. They are derived from sets of
coupled-channel equations using $T$-matrix elements constrained by
experimental meson-meson phase shifts and inelasticities, chiral
symmetry and asymptotic QCD.  We provide explicit amplitude
expressions for the decays $B^\pm \to \pi^+ \pi^- \pi^\pm$, $B \to K
\ \pi^+ \pi^-$, $B^\pm \to K^+ K^-K^\pm$, $D^+ \to \pi^- \pi^+ \pi^+
$, $D^+ \to K^- \pi^+ \ \pi^+ $, $D^0 \to K^0_S \ \pi^+\ \pi^- $, for which we have shown in previous studies
that this approach is phenomenologically successful, in addition, we provide expressions for the $D^0 \to K^0_S \ K^+ K^-$ decay. Other three-body
hadronic channels can be parametrized likewise.

\end{abstract}

\maketitle

\tableofcontents

\section{Introduction}
\label{Introduction}

Three-body hadronic decays of $B$ and $D$ mesons are a rich field for
searches on CP violation, for tests of the Standard Model and of
Quantum Chromodynamics (QCD) in particular~\cite{Ref1LHCb,Ref2LHCb,1412.4269,Ref4LHCb,Ref5LHCb}. 
Furthermore, they provide an interesting ground to study hadron physics, as strong
interaction effects, through the presence of two-body resonances and
their interferences, have an impact on weak-decay observables. In
order to obtain the latter most reliably, the meson-meson
final state interactions must be addressed using theoretical
constraints, such as unitarity, analyticity and chiral symmetry and
experimental data from processes other than $B$ and $D$ decays. However, in
Dalitz plot analyses the event distributions are often studied using
the isobar model in which the decay amplitudes are parametrized by
coherent sums of Breit-Wigner amplitudes with a background
contribution, in disagreement with the fundamental principles listed
above.  In this work, we suggest to replace these sums by parametrizations in
terms of unitary two-meson form factors, without losing contact with the description of the 
weak-interaction dynamics that governs the underlying flavor-changing process. These parametrizations 
are constructed, in part, from  results published previously~\cite{Dedonderetal2011,fkll,LZ2014,ElBennich2006,ElBennichetal09,PLB699_102,Boitoetal09,BE2009,JPD_PRD89}  
and are motivated by  the forthcoming analyses of high-statistics data sets for many three-body decay channels of $B$ and
$D$ decays, in particular by the LHC$b$ collaboration~\cite{LHCbWS}.

The theoretical amplitude expressions in Refs.~\cite{Dedonderetal2011,LZ2014,fkll,ElBennich2006,ElBennichetal09,PLB699_102,Boitoetal09,BE2009,JPD_PRD89}
from which we derive the present parametrizations are based on models of QCD factorization. The factorization beyond the 
leading approximation can be expressed as an expansion in the strong coupling, $\alpha_s$, and inverse powers of the bottom
quark mass, $m_b$, and has been applied with success to charmless nonleptonic
two-body $B$ decays~(see e.g. Ref.~\cite{Beneke2003}). Parallel analyses of three-body $B$ decays in the contexts of QCD factorization (QCDF) and perturbative QCD (pQCD) can be found in Refs.~\cite{1308.5139,1401.5514} and \cite{1502.05483, 1609.04614}, respectively.  In $D$ decays, this
factorization approach is less predictive inasmuch, as it does
not allow for a systematic improvement owing to the charm quark mass, $m_c \simeq m_b/3$,
which enhances significant corrections to the factorized results.  It
is, therefore, downgraded from an effective theory that can be
systematically improved, in the case of $B$ decays, to a phenomenological procedure, in the case
of $D$ decays.  Nevertheless, as a purely phenomenological approach, based on
the seminal work by Bauer, Stech and Wirbel~\cite{Bauer1987}, the
factorization hypothesis has been applied successfully to $D$ decays,
provided one treats Wilson coefficients as phenomenological parameters
to account for nonfactorizable corrections~\cite{Abbasetal}.

Besides a recent extension of the QCD factorization framework to nonleptonic $B$ decays into three light 
mesons~\cite{KMV2015}, no rigorous factorization theorem valid for the entire three-body phase space 
and full three light-meson Dalitz plot exists. On the other hand, three-body decays of $B$ and $D$ mesons 
clearly receive important contributions from intermediate resonances --- such as the
$\rho(770)$, $K^*(892)$ and $\phi(1020)$ --- and can therefore be considered as {\it quasi two-body decays\/}.  
One  then assumes that two of the three final-state mesons form a single state originating from a
quark-antiquark pair, which is interpreted as an intermediate
quasi-two-body final state in which case the factorization can be
applied.  Then, the three-body final state is reconstructed with the
use of two-body mesonic form factors to account for the important
hadronic final-state interactions.  For instance, in the $D^0 \to K^0_S \pi^- \pi^+ $ decay, the three-meson final state
$K^0_S \pi^+ \pi^-$ is initially preceded by the quasi-two-body pairs, $[K^0_S \pi^+]_{L} \ \pi^-$, $[K^0_S \pi^-]_{L} \ \pi^+$ and $K^0_S \ [\pi^+{\pi^-}]_{L}$, 
where two of the three mesons form a  state in an $L= S$ or $P$ wave. This framework has been successfully applied to several hadronic 
three-body $B$ and $D$ decays~\cite{MG02,fkll,ElBennich2006,ElBennichetal09,Dedonderetal2011,LZ2014,PLB699_102,Boitoetal09,BE2009,JPD_PRD89, JPD_inprogress}. 

The factorization of a nonleptonic weak $B$ decay into a quasi-two-body state can be schematically described as follows. The decays are mediated by local dimension-six 
four-quark operators $O_i (\mu)$ that form the weak effective nonrenormalizable Hamiltonian. However, depending on flavor content, spin, charge and parity symmetry of the final 
states only specific operators will contribute to a given decay. The $B$-decay amplitude into two mesons, $M_1$ and $M_2^*$ with four momenta $p_1$ and $p_2$, respectively,  can be written as,
\begin{equation}
   \langle M_1(p_1) M_2^*(p_2) | \mathcal{H}_\mathrm{eff} |  B(p_B) \rangle = \frac{G_F}{\sqrt{2}}\ V_\mathrm{CKM} \sum_i C_i(\mu) 
   \langle M_1(p_1) M_2^*(p_2) | O_i(\mu) | B(p_B) \rangle \ ,
 \label{fullamphamilton}
\end{equation}
where $p_B = p_1 + p_2$, $G_F$ is the Fermi constant, $V_\mathrm{CKM}$ is a product of Cabibbo-Kobayashi-Maskawa~(CKM) matrix elements, $C_i(\mu)$ are Wilson coefficients 
renormalized at the scale $\mu$~\cite{BurasNPB434_606} and $M_2^*(p_2)$ is the resonant quasi-two body state which decays into two lighter mesons.  The hadronic amplitude 
$\langle M_1(p_1) M_2^*(p_2) | O_i(\mu) | B(p_B) \rangle$  describes long-distance physics. In the factorization approach we henceforth employ, this amplitude is the sum of two matrix-element products,
\begin{align}
  \langle M_1(p_1) M_2^*(p_2) |  O_i(\mu) | B(p_B) \rangle & =   \Big ( \langle M_1 (p_1) |  {J}_{1}^{\nu} | B(p_B) \rangle  \langle M_2^*(p_2) | J_{2\nu} | 0 \rangle   \nonumber \\
      +  \langle M_1 (p_1) |  J_{3}^{\nu} | 0   &   \rangle  \langle M_2^*(p_2) | J_{4\nu} | B (p_B) \rangle \Big ) 
    \left [ 1 + \sum_n r_n \alpha_s^n(\mu) + \mathcal{O}  \left ( \frac{\Lambda_\mathrm{QCD}}{m_b} \right ) \right ]  \ ,
     \label{facto1}
 \end{align}
where  the strong coupling is evaluated at a scale $\mu$, $r_n$ is a combination of constant strong interaction factors,  and $\ket{0}$ is the vacuum state.
Thus, at leading order, the decay amplitudes factorize into two matrix elements with either the weak quark currents $J_1$ and $J_2$ or $J_3$ and $J_4$.
Radiative corrections  can be systematically  taken into account to a given order $\alpha_s^n (\mu)$, whereas corrections to the heavy-quark limit are of 
nonperturbative nature and therefore much less  controlled. This is in particular true for the charm quark which is neither a light nor a heavy enough
quark~\cite{ElBennich:2010ha,ElBennich:2011py,ElBennich:2012tp,Bashir:2012fs}. This fact  makes the systematic improvements of
Eq.~(\ref{facto1}), enclosed in square brackets, less reliable for $D$ decays. One should keep this limitation in mind but, for lack of a
better theoretical framework, the phenomenological approach to Eq.~(\ref{facto1})  remains a good starting point to organize the
description of $D$ decays and can be used to provide a first step beyond the isobar model.

The weak effective Hamiltonian, $\mathcal{H}_\mathrm{eff}$, in Eq.~\eqref{fullamphamilton} is given by the sum of local operators $O_i(\mu)$ multiplied by
Wilson coefficients $C_i(\mu)$ which encode the short-distance effects above the renormalization scale $\mu$. For a $\Delta B=1$
transition, for example, the
Hamiltonian is given by~\cite{Ali1998, Beneke:2001ev}
\begin{eqnarray}
\label{HeffB}
  \mathcal{H}^{\Delta B =1}_\mathrm{eff}  &=& \frac{G_F}{\sqrt{2}}   \sum_{p=u,c}  V_{pq}^* V_{pb} \left [  C_1(\mu) O_1^p(\mu) + C_2(\mu) O_2^p(\mu)    + 
                                                                             \sum_{i=3}^{10}  \hspace*{-0mm}  C_i(\mu) O_i(\mu) \right. \nonumber \\
                                                                  &+&  C_{7\gamma}(\mu) O_{7\gamma}(\mu) +  C_{8g}(\mu) O_{8g}(\mu) \Big ] +\  \mathrm{h.c.} \ ,
\end{eqnarray} 
where the quark flavor can be $q=d,s$ and $V_{ij}$ are CKM matrix elements. In the decays, the weak interaction $W$-boson exchange diagram gives rise to two 
current-current operators with different color structure owing to QCD corrections and $SU(3)$ color algebra:
\begin{eqnarray}
  O_1^p(\mu)  & = & \bar q_i \gamma^\mu (1-\gamma_5) p_i\  \bar p_j \gamma_\mu (1-\gamma_5) b_j 
  \label{q1}  \\
  O_2^p(\mu)  & = & \bar q_i \gamma^\mu (1-\gamma_5) p_j\  \bar p_j \gamma_\mu (1-\gamma_5) b_i  \ .
  \label{q2}
\end{eqnarray}
In Eqs.~\eqref{q1}  and \eqref{q2} $i,j$ are color indices and for the corresponding Wilson coefficients one has $C_1(\mu) \simeq 1 + \mathcal{O}(\alpha_s (\mu)$) and 
$C_2(\mu) \simeq \mathcal{O}(\alpha_s (\mu)$). The operators  $O_i$, $i=3-10$ stem from QCD and electroweak penguin diagrams, while $O_{7\gamma}$ and $O_{8g}$ are
electromagnetic and chromomagnetic dipole operators. The explicit tensor  structure of these operators as well as their Wilson coefficients at next-to-leading logarithms can be 
found, for example, in Ref.~\cite{Buchalla:1995vs}. With the use of an appropriate Fierz transformation and the  $SU(N_c)$ identity, 
\begin{equation}
( \bar q_i\,  p_j )\, (\bar p_j \, b_i) \  = \  2\, ( \bar q_i \, T^a_{ik}\, p_k )\,  ( \bar p_j\, T^a_{jl} \, b_l )   + \frac{1}{ N_c}\, ( \bar q_i \, p_i )\, (\bar p_j\,  b_j )   \ ,
  \label{coloridentity}  
\end{equation}
where $T^a_{ij}$ are the $SU(N_c)$ generators, the quark bilinears can be rearranged to match the flavor and color structure of the final mesons. In this transformation, the color-octet contribution in Eq.~\eqref{coloridentity}  
is commonly neglected.  The two resulting combinations of $C_1(\mu)$  and $C_2(\mu)$,
\begin{equation}
   a_1(\mu) = C_1(\mu) +\frac{1}{N_c} C_2 (\mu)  \  ,  \quad  a_2(\mu) = C_2(\mu) +\frac{1}{N_c} C_1 (\mu)  \  ,
\end{equation}
lead to ``color allowed'' and ``color suppressed" amplitudes, respectively,  which are topologically different. Typically, the Wilson coefficients are evaluated at a renormalization scale 
of the heavy quark, i.e. $\mu \simeq m_c, m_b$.

On the right-hand side of Eq.~\eqref{facto1} the two matrix-element products describe different physical processes.  Namely, the creation of a final  two-meson state from 
a $\bar qq$ pair  is described by the form factors $\langle M_2^*(p_2) | J_{2\nu} | 0 \rangle$, where $M_2^* \to M_3 M_4$ denotes resonant intermediate states in the different 
two-meson coupled channels that lead to the final three-body state. As mentioned, these form factors can be constructed such as to preserve two-body unitarity, reproduce asymptotic QCD and
are constrained by chiral symmetry at low energies. We discuss them in Appendices~\ref{AppFFs}.  In Eq.~\eqref{facto1}, the matrix element $\langle M_1 (p_1) | J_{3}^\nu | 0 \rangle$ defines 
the weak decay constant of a scalar,  pseudoscalar or vector mesons which is either well known from experiment, for instance $f_\pi$ and $f_K$, or has been evaluated with 
lattice-regularized QCD and other nonperturbative approaches.  The transition $\langle M_2^*(p_2) | J_{4\nu} | B \rangle $ of a $B$ 
meson to a strongly interacting two-meson pair  via a resonance is a complicated process and the biggest source of uncertainty in our approach.
It could be extracted experimentally  from semi-leptonic processes such as $B^0 \to K^+ \pi^- \mu^+ \mu^-$~\cite{B0semilep} or $D^0 \to K^- \pi^+ \mu^+ \mu^-$~\cite{D0semilep}. It has also been conjectured within  soft-collinear effective theory that the amplitude  can be factorized in terms of a generalized  $B$-to-two-body form factor and two-hadron light-cone distribution amplitudes~\cite{MW14}.
 In the derivation of the amplitude expressions presented here, we employ a model approximation which relates this matrix element $\langle M_2^*(p_2) | J_{4\nu} | B \rangle $ to the two-body meson form factor $\langle M_2^*(p_2) [\to M_3 M_4]| {J}_{2\nu} | 0 \rangle $.

Finally,  the transition amplitudes $\langle M_1 (p_1) | J_{1}^\nu | B \rangle $ ($=\langle M_1 (p_1) \overline {B}| J_{1}^\nu | 0 \rangle$) are parametrized by heavy-to-light transition form factors 
which are discussed in Appendix~\ref{transitionff}.  

As a definite example of the procedure outlined above, let us consider the  $D^+ \to[ K^- \pi^+]_{S,P}\  \pi^+ $ decay, where the $K^-\pi^+$
pairs are in $S$- or $P$-wave state. The matrix element given by $\langle [K^-\pi^+]_{S,P}\pi^+ | \mathcal{H}_\mathrm{eff} | D^+ \rangle$ receives 
contributions from the two amplitudes $a_1(\mu)$ and  $a_2(\mu)$ and factorizes as,
\begin{align}
\langle [ K^- \pi^+]_{S,P}\ \pi^+ | \mathcal{H}_\mathrm{eff} | D^+ \rangle & =  \frac{G_F}{\sqrt{2}}\cos^2\theta_C \Big [a_1\langle [K^-\pi^+_1]_{S,P}\vert \bar s\gamma^\nu(1-\gamma_5)c\vert D^+\rangle
                                                      \langle\pi^+_2 \vert \bar u\gamma_\nu(1-\gamma_5)d\vert 0\rangle    \nonumber    \\
                                               + \,  a_2 \langle [K^-\pi^+_1]_{S,P} & \vert \bar s\gamma^\nu(1-\gamma_5)d \vert 0\rangle
                                                    \langle \pi^+_2 \vert \bar u\gamma_\nu(1-\gamma_5)c \vert D^+\rangle \Big ] + (\pi_1^+ \leftrightarrow \pi_2^+) \, , 
\label{DKpiExample}
\end{align}
$\theta_C$ being the Cabbibo angle. The
$K\pi$ form factors appear explicitly in the matrix element $\langle
[K^-\pi^+_1]_{S,P} \vert \bar s\gamma^\nu(1-\gamma_5)d \vert
0\rangle$.  The evaluation of $\langle [K^-\pi^+_1]_{S,P}\vert \bar
s\gamma^\nu(1-\gamma_5)c \vert D^+\rangle$ is less straightforward.
However, assuming this transition to proceed through the dominant
intermediate resonances, this matrix element can also be written in
terms of the $K\pi$ form factors as shown in Ref.~\cite{MG02, BE2009}. 
 This feature  is of crucial importance to the parametrizations that we propose in this work. It is interesting to note that the calculation of a generalized three-body form factor using light-cone sum rules, in the spirit of Ref.~\cite{MW14}, also leads to the appearance of  the  two-body meson form factors~\cite{MW14n2,WZ15,SW15,Lietal16,SWZ17}.
The
other matrix elements of Eq.~(\ref{DKpiExample}) can be written in
terms of decay constants or transitions form factors that can be
extracted from semi-leptonic decays, as outlined above.  Strong phases in the mesonic final state interactions are accounted for by the 
hadronic form factors, which makes this type of description particularly suitable for the interpretation of CP asymmetries that have been observed 
in $B$ decays~\cite{ElBennichetal09,Dedonderetal2011,LZ2014, PLB699_102, Kleinetal}. Amplitude
expressions, such as in Eq.~(\ref{DKpiExample}), are used throughout
this paper as a starting point to build parametrizations based on
unitary two-body hadronic form factors.  Within this approach,
explicit forms of parametrizations for $D^+ \to K^- \pi^+ \pi^+$ and
$D^0 \to K^0_S \pi^- \pi^+$ amplitudes have already been presented in
Ref.~\cite{LHCbWS} (see p.~27 therein).

The paper is structured as follows.  In Section~\ref{3bodyBdecayAmp}
we introduce the parametrizations for three-body hadronic $B$-decay
amplitudes based on the quasi two-body factorization approaches of
Ref.~\cite{Dedonderetal2011} for $B^\pm \to \pi^+ \pi^- \pi^\pm$, of
Refs.~\cite{fkll,ElBennich2006,ElBennichetal09} for $B \to K \pi^+
\pi^-$ and of Ref.~\cite{PLB699_102} for $B^\pm \to K^+ K^-
K^\pm$. Section~\ref{3bodyDdecayAmp} applies the same procedure to $D$-decay
 amplitudes, viz. $D^+ \to \pi^+ \pi^- \pi^+
$~\cite{Boitoetal09}, $D^+ \to K^- \pi^+ \ \pi^+ $~\cite{BE2009}, $D^0
\to K^0_S \pi^+\ \pi^- $~\cite{JPD_PRD89} and $D^0 \to K^0_S
K^+\ K^-$~\cite{JPD_inprogress}.  The meson-meson and heavy-to-light
meson form factors which have been used can be found in the original
papers.  Nevertheless a short reminder on the derivations of unitary
$S$- and $P$- waves $\pi \pi$-, $\pi K$- and $K \bar K$-meson form
factors entering these parametrizations is given in
Appendix~\ref{AppFFs} together with a short review on heavy-to-light
meson form factors. We wrap up with some concluding remarks about the
merits of the proposed parametrizations in
Section~\ref{Conclusion}. The relations between the free parameters of
the different proposed parametrizations and the theoretical decay
amplitudes are presented explicitly in Appendix~\ref{relationfreeoriginal}.

\section[Parametrizations of three-body hadronic $B$-decay amplitudes]{\boldmath Parametrizations of three-body hadronic $B$-decay amplitudes}
\label{3bodyBdecayAmp}

\subsection{Amplitudes for $\boldsymbol{B^\pm \to \pi^\pm \pi^+ \pi^-}$ } 
\label{QtbAmpBppp}

The contributions of pion-pion interactions to CP violating phases in $B^\pm\to \pi^\pm \pi^\mp \pi^\pm$ decays have been studied~\cite{Dedonderetal2011} within the quasi two-body 
factorization approach discussed in the introduction.\footnote{During the preparation of this manuscript Ref.~\cite{Kleinetal} has appeared. Their treatment is very similar to the one we describe here.} 
The amplitudes were derived as matrix elements of the weak effective Hamiltonian 
given by Eq.~\eqref{HeffB} with $q\equiv d$. The $\pi \pi$ effective mass distributions of the  $B^\pm \to \pi^\pm \pi^+ \pi^-$ data~\cite{Aubert:2009} are well reproduced for 
an invariant mass, $m_{\pi^+ \pi^-}\lesssim$~1.64~GeV~\cite{Dedonderetal2011}. To parametrize the amplitudes of $B^\pm\to \pi^\pm [\pi^+\pi^-]_{S,P}$
we label the momenta of the decay  as $B^\pm(p_B)  \to \pi^\pm (p_1) \pi^+(p_2) \pi^-(p_3)$, where $p_B$,   the $B^\pm$ meson momentum, satisfies $p_B=p_1+p_2+p_3$. 
The amplitudes must be symmetrized by exchanging the $\pi^+(p_2) \pi^-(p_3)$ and $\pi^-(p_1) \pi^+(p_2)$ pairs in case of a $B^-$ decay or  equivalently the 
$\pi^+(p_2) \pi^-(p_3)$ and $\pi^+(p_1) \pi^-(p_3)$ pairs in case of a $B^+$ decay. Defining the invariants, $s_{ij}= (p_i + p_j)^2$ (for $i\neq
j$), with $s_{12}+ s_{13}+s_{23}= m_B^2+3 m_\pi^2$, the  interacting pairs of pions  in a relative $S$ or $P$ wave are described by 
$s_{12}$ or $s_{23}$, in the case of a $B^-$ decay, and by $s_{13}$ and $s_{23}$ in the case of  a $B^+$ decay.

The symmetrized amplitude (see also Eq.~(21) in Ref.~\cite{Dedonderetal2011}) for the $B^- \to \pi^- [\pi^+\pi^-]_{S,P}$ decay reads, 
\begin{eqnarray}
\label{Mmoinsinv}
\hspace*{-3mm}
\mathcal{A}^-_\mathrm{sym}(s_{12}, s_{23})  = \frac{1}{\sqrt{2}}\left [\mathcal{A}^-_S(s_{12})+
  \mathcal{A}^-_S(s_{23}) + (s_{13}-s_{23}) \mathcal{A}^-_P(s_{12})  + (s_{13}-s_{12})\mathcal{A}^-_P(s_{23})  
\right ],
\end{eqnarray}
and an analogous amplitude holds for the $B^+ \to \pi^+ [\pi^-\pi^+]_{S,P}$ decay . The amplitudes $\mathcal{A}^-_{S,P}(s_{ij})$, $ij= 12$ or $23$, 
given by Eqs.~(22) and  (23) of Ref.~\cite{Dedonderetal2011}, can be parametrized in terms of four complex parameters,  $a_{1,2}^{S,P}$ as,\footnote{In a fit to a Dalitz plot there is always a global phase that cannot be observed. Therefore, the phase of one of the complex parameters can be set to zero. This is also valid for the other channels discussed in the remainder of this paper.}
\begin{eqnarray}
\label{MSsij}
\mathcal{A}^-_S(s_{ij})&=& \Big  [ a_1^S\left(M_{B}^2-s_{ij}\right) + a_2^S F_0^{B\pi}(s_{ij}) \Big ] F_{0n}^{\pi \pi}(s_{ij}),\\
\label{MPsij}
\mathcal{A}^-_P(s_{ij})&=&\Big [a_1^P + a_2^P F_1^{B\pi}(s_{ij})  \Big ] F_1^{\pi \pi}(s_{ij}),
%
\end{eqnarray}
where $M_B$ is the charged $B$-meson mass. As done in the BABAR
collaboration analysis~\cite{Aubert:2009} and in
Ref.~\cite{Dedonderetal2011}, a contribution from the $f_2(1270)$ resonance
can be accounted for by a Breit-Wigner line shape in a $D$-wave
amplitude of the $\pi^+ \pi^-$ pair.  The $B \to \pi$ scalar and
vector transition form factors $F_{0,1}^{B\pi}(s)$ in
Eqs.~(\ref{MSsij}) and (\ref{MPsij}) are discussed in
Appendix~\ref{transitionff}.  The $\pi\pi$ $S$-wave amplitude
$\mathcal{A}^-_S(s_{ij})$ includes via the non-strange scalar form
factor $F_{0n}^{\pi \pi}(s_{ij})$ the contributions of the scalar
$f_0(500)$, $f_0(980)$ and $f_0(1400)$ resonances. In a Dalitz-plot
analysis, one can use, for example, the  pion scalar form factor
derived in Refs.~\cite{Dedonderetal2011} and \cite{Moussallam_2000}. More
details are given in Appendix~\ref{Fpp}.

The $P$-wave amplitude $\mathcal{A}^-_P(s_{ij})$, proportional to the
pion vector form factor $F_1^{\pi \pi}(s_{ij})$, contains the
$\rho(770)^0$, $\rho(1450)$ and $\rho(1700)$ contributions. In
Ref.~\cite{Dedonderetal2011}, the $(\pi\pi)_P$ form factor was extracted
from the Belle Collaboration analysis of $\tau^-\to\pi^-\pi^0\nu_\tau$
decay data~\cite{Fujikawa_PRD78_072006}. Alternatively, one can employ
the unitary parametrization of Ref.~\cite{Hanhart} which fits the
$(\pi\pi)_P$-wave phase shifts and inelasticities, the $e^+ e^- \to \pi^+
\pi^-$ data and the $\tau^-\to\pi^-\pi^0\nu_\tau$ decay data, as done
in the $D^0 \to K^0_S \pi^+ \pi^-$ Dalitz plot fit of
Ref.~\cite{JPD_PRD89}; see Appendix~\ref{Fpp}.

Setting the phase of $a_1^P$ in $\mathcal{A}^-_P(s_{ij})$ to zero
yields a total of 7 real parameters to be fitted. The fully
symmetrized CP-conjugate $B^+ \to \pi^+ \pi^- \pi^+ $ decay
amplitude is given by expressions similar to
Eqs.~(\ref{Mmoinsinv})--(\ref{MPsij}) with again 7 free real
parameters.  The reproduction of the Dalitz-plot data over the full
phase space, in particular for the high invariant mass regions, might
require some adjustment of the $\pi \pi$ form factors. The addition of
further phenomenological amplitudes that represent contributions of
higher $\pi \pi$-interacting waves and possible three-body
rescattering terms may be necessary.

\subsection{Amplitudes for $\boldsymbol{ B \to K \pi^+ \pi^-} $}
\label{QtbAmpBKpp}

The amplitude is based on the weak effective Hamiltonian in Eq.~(\ref{HeffB}) with $q\equiv s$. The momenta are labelled as $B(p_B)\to K(p_1) \pi^+(p_2)\pi^ -(p_3)$, 
with $s_{12}=(p_1+p_2)^ 2$, $s_{13}=(p_1+p_3)^ 2$,  $s_{23}=(p_2+p_3)^ 2$ and  $s_{12}+ s_{13}+s_{23}= m_B^2+m_K^2+2 m_\pi^2$ .

\subsubsection{Parametrization of  the $B \to K [\pi^\pm \pi^\mp]_{S,P}$ amplitude}
\label{pipiSPwavepKAmp}

The isoscalar $S$-wave $\pi^+\pi^-$ final state interactions in $B\to
K\pi^+\pi^-$ decays were studied in Ref.~\cite{fkll} in the quasi
two-body factorization approach with an extension in
Ref.~\cite{ElBennich2006} to include the $\pi^+\pi^-$ isovector $P$
wave. These studies reproduce very well the Belle and BABAR data in an
effective $\pi \pi$ mass range up to about 1.2~GeV. Following Eq.~(1) of Ref.~\cite{fkll} the $B \to K [\pi^+ \pi^-]_{S}$
decay amplitude can be parametrized in terms of three complex
parameters, $b_i^S, i=1,2,3$, for the different charges $B= B^\pm, K =
K^\pm$ and $B= B^0 (\bar B^0), K = K^0(\bar K^0)$ or $K^0_S$,
\begin{equation}
\label{paracs}
  \mathcal{A}_S(s_{23})  \equiv  \langle K \ [\pi^+\pi^-]_S \vert \mathcal{H}_{\rm eff} \vert B \rangle = 
   b_1^S \left (M_B^2-s_{23} \right ) F_{0n}^{\pi\pi} (s_{23})  + \left ( b_2^S F_0^{B K}(s_{23})+b_3^S \right ) F_{0s}^{\pi\pi}  (s_{23}).
\end{equation}
For the scalar-isoscalar strange form factor $ F_{0s}^{\pi\pi} (s)$ in Eq.~(\ref{paracs}), one can employ its numerical expression
given in Ref.~\cite{Dedonderetal2011} or that in Ref.~\cite{Moussallam_2000}~(see Appendix~\ref{SFpp}). Parametrizations for the $B \to K$ 
scalar transition form factors $F_{0}^{BK}(s)$ in Eq.~\eqref{paracs} are reviewed in Appendix~\ref{transitionff}.

The amplitude for $B \to K [\pi^+\pi^-]_{P}$ decays can be written in
terms of the complex parameter $b_1^P$ as,
\begin{equation}
\label{paracp}
 \mathcal{A}_P(s_{12}, s_{13}, s_{23})  \equiv \langle K \ [\pi^+\pi^-]_P \vert \mathcal{ H}_{\rm eff} \vert B  \rangle=  b_1^P \left (s_{13} - s_{12}\right ) F_1^{\pi \pi}(s_{23}).
\end{equation}
In Ref.~\cite{ElBennich2006}, the pion vector form factor, $
F_{1}^{\pi\pi} (s)$, is approximated by a Breit-Wigner form. However,
we recommend the use of the unitary vector form factor derived in
Ref.~\cite{Hanhart} described in Appendix~\ref{VFpp}. We stress that
the $b_i^S$ in Eq.~(\ref{paracs}) and $b_1^P$ in Eq.~(\ref{paracp})
represent different parameters for each charge state.

As in the $B^\pm \to \pi^+ \pi^- \pi^\pm$ case (see
Section~\ref{QtbAmpBppp}), addition of the $[\pi^+ \pi^-]$ $D$-wave
contribution, parametrized in terms of the $f_2(1270)$ resonance, is
required; higher invariant-mass phenomenological amplitudes may also
be necessary.

\subsubsection{Parametrization of  the $B \to [K \pi^\pm]_{S,P} \pi^\mp$ amplitude}
\label{SwavepiKAmp}
 
A parametrization of the $B \to [K \pi^\pm]_{S} \pi^\mp$ channel was
introduced in Ref.~\cite{ElBennichetal09} (see Eq.~(68) therein),
where in the center-of-mass of the $K\pi$ pair the $S$-wave amplitude
in case of the $B^- \to [K^-\pi^+]_S\pi^-$ decay can be represented by,
\begin{equation}
\label{MSparam}
   \mathcal{A}_S (s_{12}) \equiv \langle \pi^- \ [ K^- \pi^+]_S \vert \mathcal{H}_\mathrm{eff}\vert B^- \rangle = 
    \big ( c_1^S + c_2^S s_{12} \big )\, \frac {F_0^{ B \pi }  (s_{12} ) F_0^{K \pi} (s_{12} )}{s_{12}} \ ,
\end{equation}
which follows from Eq.~(10) of Ref.~\cite{ElBennichetal09}. In
Eq.~(\ref{MSparam}), $s_{12}$ is the invariant mass squared of the interacting
$K^-\pi^+$ pair, whereas for $B^+$ and $\bar B^0$ decays the kinematic
variable is $s_{13}$.  The complex parameters, $c_1^S$ and $c_2^S$,
can be determined through the Dalitz-plot analysis for each given
charge state.  We note that the isolated $K_0^*(1430)$ resonance
contribution can be obtained by replacing, once the parameters $c_1^S$
and $c_2^S$ are determined, $F_0^{K\pi} (s )$ by its pole part
$F_0^\mathrm{pole} (s)$ given in Eqs.~(45)$-$(47) of
Ref.~\cite{ElBennichetal09}.

Following the momentum conventions of the $S$-wave above, the  $K \pi$ $P$-wave amplitude of the  $B^- \to [K^- \pi^+]_P\pi^-$ decays can be parametrized as,
\begin{eqnarray}
\label{MPparam}
\mathcal{A}_P (s_{12}, s_{23}) & \equiv & \langle \pi^- \ [ K^- \pi^+ ]_P  \vert \mathcal{H}_\mathrm{eff}\vert B^- \rangle 
          \nonumber \\ 
        & = &  c_1^P  \left( s_{13} - s_{23} - (M_B^2-m_\pi^2)\ \dfrac{m_K^2-m_\pi^2}{s_{12}}  \right  )  F_1^{B \pi } ( s_{12} ) F_1^{K\pi} (s_{12}) \ .
\end{eqnarray}
The parametrizations for the transition form factors, $F_{0(1)}^{ B\pi
}(s)$, are discussed in Appendix~\ref{transitionff} and those for the
$K\pi$ scalar and vector form factors $F_{0(1)}^{K \pi } (s)$ in
Appendix~\ref{Kp}.
     
The then available Belle and BABAR data were well reproduced in the
$\mathcal{A}_{S,P}$ amplitude analysis of Ref.~\cite{ElBennichetal09}
in a $m_{K\pi}$ range from threshold up to 1.8~GeV. Within the
factorization approximation there is no contribution from $[K\pi]$
partial waves, $l\geq 2$, thus one expects the $[K\pi]$ $D$-wave
contribution to be small. However, in order to analyze the Dalitz-plot
data over the full energy ranges, additional phenomenological
amplitudes are required.

\subsection{Amplitudes for  $\boldsymbol{ B^\pm \to K^\pm K^+ K^- }$}
\label{QtbAmpBKKK}

The weak effective Hamiltonian that describes this decay channel is
given by Eq.~(\ref{HeffB}) with $q\equiv s$.  Explicit factorized
expressions of $B^- \to K^- K^+ K^-$ amplitudes can be found in
Appendix~A of Ref.~\cite{Cheng0704.1049}.  The effective invariant
$K^+ K^-$ mass distributions of the decays $B^\pm \to K^+ K^-
K^\pm$~\cite{Belle,BABAR} up to $1.8$~GeV were shown to be well
reproduced in the factorization approach of Ref.~\cite{PLB699_102},
where the $B^\pm (p_B) \to K^\pm(p_1) K^+(p_2) K^-(p_3)$ amplitudes
were derived for interacting $K^+ K^-$ pairs in a relative $S$ or $P$
state.  The symmetrized term for a $B^-$ decay is obtained by exchange
of the $K^+(p_2) K^-(p_3)$ pair with the $K^-(p_1) K^+(p_2)$ one (or
exchanging the $K^+(p_2) K^-(p_3)$ and $K^+(p_1) K^-(p_3)$ pairs in
case of a $B^+$ meson) and is added to the amplitude.  The totally
symmetrized amplitude using the Lorentz invariants $s_{ij}= (p_i +
p_j)^2$ for $i\neq j$ is given by,
\begin{eqnarray}
\label{Amoinsinv}
\hspace*{-5mm}
      \mathcal{A}^-(s_{12}, s_{13}, s_{23})& = & \frac{1}{\sqrt{2}}\big [ \mathcal{A}^-_S(s_{12})+  \mathcal{A}^-_S(s_{23}) + \mathcal{A}^-_P(s_{12}) (s_{13}-s_{23})
   + \mathcal{A}^-_P(s_{23}) (s_{13}-s_{12}) \big ] ,
\end{eqnarray}
with $s_{12}+ s_{13}+s_{23}= m_B^2+3 m_K^2$.

Following Eqs.~(2) and (3) of Ref.~\cite{PLB699_102}, the amplitudes $\mathcal{A}^-_{S,P}(s_{ij})$, $ij= 12$ or $23$, can be also written in terms of six complex 
parameters  $d_{1,2}^S$ and  $d_{1,2,3,4}^P$
\begin{eqnarray}
\label{ASsij} 
\mathcal{A}^-_S (s_{ij})&=&  d_1^S \left(M_{B}^2-s_{ij}\right) F^{KK}_{0n} (s_{ij})  +  d_2^S F_0^{BK}(s_{ij})  F_{0s}^{KK} (s_{ij}) \ , \\
\label{APsij}
 \mathcal{A}^-_P (s_{ij})&=& d_1^P F^{KK}_{1u}(s_{ij}) 
                         + F^{BK}_1(s_{ij}) \left [d_2^P F^{KK}_{1u} (s_{ij}) + d_3^P  F^{KK}_{1d} (s_{ij})+ d_4^P F^{KK}_{1s} (s_{ij}) \right ].
\end{eqnarray}
where the $B \to K$ scalar and vector transition form factors $F_{0,1}^{BK}(s)$ in Eqs.~(\ref{ASsij}) and (\ref{APsij}) are discussed  in Appendix~\ref{transitionff} and
the scalar and vector form factors $F^{KK}_{0n}, F^{KK}_{0s}, F^{KK}_{1u}$, $F^{KK}_{1d}$ and $F^{KK}_{1s}$ are introduced in Appendix~\ref{KK}. 

Because of its small branching fraction to $K \bar K$ (4.6\%) the $f_2(1270)$ contribution was not introduced in Ref.~\cite{PLB699_102}.
However, in their study of the Dalitz-plot dependence of CP asymmetry, the authors of Ref.~\cite{LZ2014} have included it and an amplitude analysis of the full Dalitz
plot should also add it together with a phenomenological term representing
the high invariant-mass contributions.


\section{Parametrizations of three-body hadronic $\boldsymbol{D}$-decay amplitudes}
\label{3bodyDdecayAmp}

\subsection{Amplitudes for  $\boldsymbol{ D^+ \to \pi^+ \pi^- \pi^+ }$}
\label{QtbAmpDppp}

The decay $\Dppp$ is a Cabibbo suppressed mode governed by the quark-level transition $c\to du\bar d$. The leading contribution to
the amplitude arises from the current-current operators and is proportional to $V_{cd}V_{ud}^*$, which is $\mathcal{O}(\lambda)$ in
Wolfenstein parametrization, with $\lambda = 0.2257$.  At next-to-leading order (NLO) in QCD, penguin operators contribute to the 
decay amplitude. However, CKM unitarity implies that those come with a coefficient $V_{cb}V_{ub}^*$, which is $\mathcal{O}(\lambda^5)$. 
It is therefore safe to neglect the penguin contributions. The dominant contributions to the effective Hamiltonian therefore are, 
\begin{equation}
\label{effHDpppp}
{\cal H}_{\rm eff}=\frac{G_F}{\sqrt{2}} V_{cd}V_{ud}^\ast\, \big  [C_1(\mu)O_1+C_2(\mu)O_2 \big ]\ + \ \mathrm{h.c.}  \ ,
\end{equation}
where the two four-quark operators read,
\bea
\label{effO12D3pi}
O_1 &=&  \big [\bar d_i\gamma^\nu (1-\gamma_5) c_i][\bar u_j\gamma_\nu (1-\gamma_5) d_j  \big ] \ ,  \\
O_2 &=&  \big [\bar u_i\gamma^\nu (1-\gamma_5) c_j][\bar d_j\gamma_\nu (1-\gamma_5) d_i  \big ]  \ .
\eea
In the description of $D$ decays, however, because of the limitations discussed in the Introduction,  the Wilson coefficients
depart from their calculated values due to  nonfactorizable
corrections.  In the spirit of Ref.~\cite{Abbasetal}, it is safe to assume they
can be complex numbers and that the corrections will depend on whether
the  $\pi\pi$ pair  is in $S$- or $P$-wave state (the same applies to the $K\pi$ pairs in the next section). Our parametrizations below encompass this assumption.

The parametrization we propose here is chiefly based on the work of Ref.~\cite{Boitoetal09}.  The crucial dynamical ingredient
to describe the two-body hadronic final state interactions are the scalar and vector $\pi\pi$ form factors. The description of the decay proceeds in full
analogy to that outlined in the introduction for $\DKpp$.  Within this framework, the main difference between these two decays is that 
the relevant two-pion matrix element, namely, $\sand{\pi^-\pi^+}{\bar d\gamma^\nu(1-\gamma_5)d}{0}$ is proportional to the $\pi\pi$ vector form factor 
only --- no scalar contribution appears because the pseudo-scalars involved have the same mass. This implies that the $\pi^+\pi^-$ $S$-wave in the 
decay $\Dppp$ receives no contribution from the diagram proportional to $a_2(\mu)$. The decay amplitude for $S$- and $P$-wave $\pi^+\pi^-$ pairs can be written as
\beq
\sand{\pi^+[\pi^-\pi^+]_{S,P}}{\mathcal{H}_{\rm eff}}{D^+} = \mathcal{A}_{S}^{+}+ \mathcal{A}_P^{+}    ,\label{eq:DpppSandP}
\eeq
where $\mathcal{A}_S^+$, $\mathcal{A}_P^+$  are respectively the $S$- and  $P$-wave $\pi^+\pi^-$ amplitudes. 
The $S$-wave  is dominated by the intermediate scalar-isoscalar resonances $f_0(500)$ and $f_0(980)$ while the $P$-wave  is largely dominated by the $\rho(770)^0$.

We label the four-momenta as $D^+(p_D) \to \pi^+(p_1)\pi^-(p_2)\pi^+(p_3)$,
and define the following invariant masses squared $s_{12} = (p_1+p_2)^2$, $s_{23} = (p_2+p_3)^2$, and  $s_{13} = (p_1+p_3)^2$, with $s_{12}+ s_{13}+s_{23}= m_D^2+3 m_\pi^2$.
Resonances occur in the $\pi^+\pi^-$ states described in terms of $s_{12}$ and  $s_{23}$ invariants. With these definitions the amplitudes $\mathcal{A}_{S,P}^+$ of Eq.~(\ref{eq:DpppSandP}) 
can be parametrized with three complex  parameters, $e_1^{S}$ and $e_{1,2}^P$, as:
\begin{align}
\label{AD3piS}
\mathcal{A}_{S}^{+} (s_{12}, s_{23})&=  e_1^S (m^2_D - s_{12}) F^{\pi\pi}_{0n}(s_{12}) + (s_{12}\leftrightarrow s_{23}), \\
\mathcal{A}_{P}^{+} (s_{12}, s_{13}, s_{23})&=   \left[  e_1^{P} +  e_{2}^{P}  F_1^{D\pi}(s_{12})   \right]  (s_{23} - s_{13})F_1^{\pi\pi}(s_{12})  + (s_{12}\leftrightarrow s_{23}) \ .
\label{AD3piP}
\end{align}
We are implicitly assuming that non-factorizable corrections depend on
the spin of the $\pi^+\pi^-$ pair and can be absorbed in the parameters
$e_i^L$.  In this parametrization, the two-body $\pi^+\pi^-$
interactions are fully taken into account by the scalar and vector
$\pi\pi$ form factors, $F_{0n}^{\pi\pi}$ and $F_1^{\pi\pi}(s)$,
respectively, which are detailed in Appendix~\ref{Fpp}. The vector $D\to \pi$ transition form factor, $F_1^{D\pi}(s)$, in Eq.~(\ref{AD3piP}) is discussed in Appendix~\ref{transitionff}. We observe that the $D$-wave resonance contribution, arising from the $f_2(1270)$, is
sizeable (with fit fractions of about $20\%$~\cite{PDG,E791Dp3pi}) and could be included in data analyses through usual isobar model
expressions. Finally, in one of the models employed by the CLEO collaboration~\cite{CLEODp3pi}, some evidence for a contribution from isospin-2
$\pi^+\pi^+$ interactions is presented, which may have to be included in a realistic analysis.

\subsection{Amplitudes for $\boldsymbol{  D^+ \to K^- \pi^+  \pi^+ }$} 
\label{QtbAmpDKpp}

The $\DKpp$ decay is  Cabibbo allowed, governed by the quark-level transition  $c\to su \bar d$. Since four different quark flavors intervene, the effective Hamiltonian
for this processes does not include penguin-type operators. At NLO in QCD, there are only two operators  to be considered,
\begin{equation}
\label{effHDpKpp}
{\cal H}_{\rm eff}=\frac{G_F}{\sqrt{2}}V_{cs}V_{ud}^\ast \big [ C_1(\mu)O_1+C_2(\mu)O_2 \big ] \ +\ \mathrm{h.c.} \ ,
\end{equation}
where the relevant four-quark operators are
\bea
\label{effO12DKpp}
O_1 &=& \big [\bar c_i\gamma^\nu (1-\gamma_5) s_i][\bar d_j\gamma_\nu (1-\gamma_5) u_j \big ] ,\\
O_2 &=& \big [\bar c_i\gamma^\nu (1-\gamma_5) s_j][\bar d_j\gamma_\nu (1-\gamma_5) u_i \big ] .
\eea

In Ref.~\cite{BE2009} the $K\pi$ $S$- and $P$-wave amplitudes in this
decay were written in terms of the scalar and vector $K\pi$ form
factors, $F_{0,1}^{K\pi}(s)$.  We use these results as the basis for
our suggested parametrization.  We label the momenta as $D^+(p_D) \to
\pi^+(p_1) \pi^+(p_2)K^-(p_3)$ and define the following invariant
masses squared of the final state, $s_{13}=(p_{1}+p_{3})^2$,
$s_{23}=(p_{2}+p_{3})^2$ and $ s_{12}=(p_{1}+p_{2})^2$, with $s_{12}+
s_{13}+s_{23}= m_D^2+m_K^2+2 m_\pi^2$.  Thus, the $S$- and $P$-wave
amplitudes for $D^+ \to [K^- \pi^+]_{S,P}\ \pi^+$ are,
\beq
\sand{[K^-\pi^+]_{S,P}\pi^+}{{\cal H}_{\rm eff}}{D^+} =  \mathcal{A}_S^+  + \mathcal{A}_P^+    .
\eeq
Contribution from  $D$-wave resonances  are known to be rather small in this decay~\cite{PDG}. The $S$- and $P$-wave 
amplitudes can be parametrized with complex parameters, $f_{1,2}^{S}$ and $f_{1,2}^{P}$, as follows:
\begin{align}
\label{ASPD+1} 
\mathcal{A}^+_S(s_{13}, s_{23}) & =  \left [ f_1^{S}\, (m_D^2 -s_{13}) + f_2^S \, \frac{F_0^{D\pi}(s_{13})}{s_{13}}   \right] F_0^{K\pi}(s_{13}) +(s_{13} \leftrightarrow s_{23}),  \\
\mathcal{A}^+_P(s_{12}, s_{13}, s_{23})&  =  \left[ f_1^P\,  \Omega(s_{13},s_{23}) + f_2^P \left( \frac{s_{23} - s_{12}}{\Delta^2}    - \frac{1}{s_{13} } \right ) 
  F_1^{D\pi} (s_{13})  \right]F_1^{K\pi}(s_{13})+(s_{13} \leftrightarrow s_{23}),
  \label{ASPD+2} 
\end{align}
with $\Omega(s_{13},s_{23}) = s_{23}-s_{12}-{\Delta}^2/s_{13}$ and ${\Delta}^2=({m_{K}}^2 -m_\pi^2)(m^2_{D^+}-m^2_\pi)$. We assume that nonfactorizable corrections
are absorbed in the complex  parameters, $f_{1,2}^{S}$ and $f_{1,2}^{P}$, to be fitted to the data.  

The parametrization we introduce in Eqs.~\eqref{ASPD+1} and
\eqref{ASPD+2} makes the emergence of the scalar and vector $K\pi$
form factors, $F_{0,1}^{K\pi}(s)$, explicit. They are discussed in
more detail in Appendix~\ref{Kp}.  The scalar and vector $D\pi$
transition form factors also appear in Eqs.~(\ref{ASPD+1}) and
(\ref{ASPD+2}). Their variation with energy for the physical values of
$s$ is not significant, but they affect the shape of the amplitudes
close to the edges of the Dalitz plot. Possible parametrizations are
discussed in Appendix~\ref{transitionff}.

A version of the above description put forward here has been employed
successfully in Ref.~\cite{BE2009}. Additional contributions,
e.g. with higher angular momentum or the isospin-2 $\pi^+\pi^+$
interactions\footnote{Most experimental analyses agree this
  contribution is negligible~\cite{E791DKpipi,FOCUSDKpipi} with the
  exception of the CLEO collaboration analysis where a fit fraction of
  $\sim 20\%$ is attributed to the $\pi^+\pi^+$
  interactions~\cite{CLEODKpipi}.}, are small in this process.  In a
realistic high-statistics Dalitz plot analysis, however, they may be
required and would have to be included in the signal function through
usual isobar model expressions, for instance.

A final ingredient that is not present in our
  parametrizations are the genuine three-body hadronic final state
  interactions, which are most often neglected in experimental
  analyses. Their treatment is somewhat involved and only recently
  this problem started to be dealt with. An approach based on Feynman
  diagrams from effective lagrangians was introduced in
  Refs.~\cite{Magetal2011,MR15} precisely in the case of $D^+\to
  K^-\pi^+ \pi^+$. Alternatively, a description based on a dispersive treatment
 introduced in \cite{NKS12} for $(\omega/\phi)\to
  \pi\pi\pi$ decays has been applied to $D\to K \pi\pi$ decays in
  Refs.~\cite{NK15,NK17}. Finally, a coupled channel description
  including three-body scattering was performed in
  Ref.~\cite{Nakamura}. These treatments do not allow for a simple
  parametrization of the type we advocate here with the goal of
  replacing isobar model expressions. These three-body effects, if
  important, should show as deviations from our description and
  represent a refinement to the amplitudes discussed here that should be addressed in the future.

\subsection{Amplitudes  for $\boldsymbol{  D^0 \to K^0_S \pi^+\ \pi^- }$}
\label{QtbAmpD0K0pp}

The decay $D^0 \to K^0_S \pi^- \pi^+ $ was treated within the
framework of quasi-two body factorization in Ref.~\cite{JPD_PRD89}.  A
good reproduction of the Belle Dalitz plot density distributions~\cite{A.Poluektov_PRD81_112002_Belle} was
obtained and so were the distributions produced by the BABAR
model.\footnote{The model is built from a fit to the BABAR Dalitz plot
  data; see Ref.~\cite{JPD_PRD89}.}  The parametrizations that follow
in the next subsections are based on the quasi-two-body amplitudes
derived in this study. The Hamiltonian that describes this decay
channel is similar to that of Eq.~(\ref{effHDpKpp}), but besides a
Cabibbo favored term proportional to $V^*_{cs}V_{ud}$ there is also a
doubly Cabibbo suppressed contribution proportional to
$V^*_{cd}V_{us}$. The momenta are labelled as $D^ 0(p_D)\to K^0_S(p_1)
\pi^-(p_2) \pi^+(p_3)$ where the kinematic configuration is defined by
$s_{12}=(p_1+p_2)^2$, $s_{13} =(p_1+p_3)^2$ and $s_{23}=(p_2+p_3)^2$,
with $s_{12}+ s_{13}+s_{23}= m_{D^0}^2+m_{K^0}^2+2 m_\pi^2$. We start
with the parametrization of the amplitude for the interacting $K^
0_S\pi^ -$ in an $S$- or $P$-wave state.

\subsubsection{Parametrization of the $D^0 \to [K^0_S \pi^-]_{S,P}\ \pi^+ $ amplitudes}
\label{paramD0Kp-p+}

The following parametrizations are derived from Eqs.~(66) and (68) of Ref.~\cite{JPD_PRD89}.
In terms of three complex parameters $g_{1,2}^ S$ and $g_1^ P$, the parametrized amplitudes read
\begin{eqnarray}
\label{ASD0}
\mathcal{A}^{0}_{S,-}(s_{12})&=& \left( g_{1}^ S+g_{2}^S s_{12} \right) F_0^{K\pi}(s_{12}),  \\
\mathcal{A}^{0}_{P,-}(s_{12}, s_{13}, s_{23})&=& g_{1}^ P\left( s_{23} - s_{13} +\frac{\Delta_0^2}{s_{12}}\right)  F_1^{K\pi}(s_{12}),
\label{APD0}
\end{eqnarray}
with ${\Delta_0}^2=({m_{K^0}}^2 -m_\pi^2)(m^2_{D^0}-m^2_\pi)$.  The
$\pi K$ $S$-wave $\mathcal{A}^{0}_{S,-}$ amplitude includes the
contribution of the scalar $K^*_0(800)^-$ and $K^*_0(1430)^-$
resonances and the $P$-wave $\mathcal{A}^{0}_{P,-}$ that of the vector
$K^*(892)^-$.  Despite its small fit fraction, the $\pi K$ $D$-wave
$D^0 \to [K^0_S \pi^-]_{D}\ \pi^+ $ amplitude plays an important role
through interference.  The contribution of the tensor $K^{*-}_2(1430)$
resonance can be parametrized by a relativistic Breit-Wigner formula, with a
magnitude and phase that should be obtained through a fit to the data,
as done successfully in Ref.~\cite{JPD_PRD89}. This component should
be added to the $S$ and $P$ wave amplitudes parametrized above.

\subsubsection{Parametrization of the $D^0 \to [K^0_S \pi^+]_{S,P}\ \pi^- $ amplitudes}
\label{paramD0Kp+p-}

Likewise, the decay amplitudes for $D^0 \to [K^0_S \pi^+]_{S,P}\ \pi^- $ are given in Ref.~\cite{JPD_PRD89}  [see Eqs.~(84) and (85)]
and can be parametrized as, 
\begin{eqnarray}
\label{A8SPD0}
\mathcal{A}^{0}_{S,+}(s_{13})&=& \left[ g_3^ S(m_\pi^2-s_{13})+g_4^ S\ \frac{\Delta_0^2}{s_{13}} F_0^{D\pi}(s_{13}) \right] F_0^{K\pi}(s_{13}) \ , \\
\label{A9SPD0}
\mathcal{A}^{0}_{P,+}(s_{12}, s_{13}, s_{23})&=&\left[ g_2^ P+ g_3^ P F_1^{D\pi}(s_{13})\right] \left( s_{23}-s_{12} +\frac{\Delta_0^2}{s_{13}}\right)   F_1^{K\pi}(s_{13}) \ .
\end{eqnarray}

The  $\pi K$ $S$-wave $\mathcal{A}^{0}_{S,+}$ amplitude includes the contribution of the  scalar $K^*_0(800)^+$ and $K^*_0(1430)^+$ resonances, 
the $P$-wave $\mathcal{A}^{0}_{P,+}$ that of  the vector $K^*(892)^+$. The contribution from
the $D$-wave, that stems mainly from the $K^*_2(1430)^+$, could be parametrized by the usual Breit-Wigner expressions.

\subsubsection{Parametrization of the $D^0 \to K^0_S [\pi^+ \pi^-]_{S,P}$ amplitudes}
\label{paramD0K0p+p-}

The weak $D^0 \to K^0_S [\pi^+ \pi^-]_{S,P}$  decay amplitudes, following
Ref.~\cite{JPD_PRD89}, can be parametrized, using the 
same momentum definition as before, as: 
\begin{eqnarray}
\label{A2SPD0}
\mathcal{A}^{0}_{S,0}(s_{23})&=& \left( g_5^S+g_6^S s_{23}\right) F_{0n}^{\pi\pi}(s_{23})\ , \\
\label{A4SPD0}
\mathcal{A}^{0}_{P,0}(s_{12}, s_{13}, s_{23})&=& \left( s_{12} - s_{13} \right)   \left [g_{4}^ P F_1^{\pi\pi}(s_{23})+g_5^P F_1^{\omega}(s_{23}) \right ] \ .
\end{eqnarray}
The $\pi\pi$ $S$-wave $\mathcal{A}^{0}_{S,0}$ amplitude includes the
contributions of the scalar $ f_0(500)$ (or $\sigma), f_0(980)$ and
$f_0(1400)$ resonances. The $[\pi^+\pi^-]_P$ pair can originate from
the $\omega$ resonance through isospin violation. This introduces
a term proportional to the vector form factor $F_1^\omega(s_{23})$ in
Eq.~(\ref{A4SPD0}).\footnote{In Eq.~(71) of Ref.~\cite{JPD_PRD89} this term was explicitly written as $F_1^\omega(s_{23})= m_\omega^2/(m_\omega^2-s_{23}-i m_\omega \Gamma_\omega)$.}
The effects of the vector $\rho(770)^0$ and
$\omega(782)$ resonances are included in the $P$-wave amplitudes,
$\mathcal{A}^{0}_{P,0}$, which also contain the contribution of the
$\rho(1450)^0$ and $\rho(1700)^0$; see details in
Appendix~\ref{Fpp}. The $D$-wave is dominated by the $f_2(1270)$
tensor meson and must be included in a realistic amplitude. In
Ref.~\cite{JPD_PRD89} it was parametrized by the usual relativistic
Breit-Wigner line shape.

The full decay  amplitude is thus given by,
\begin{equation}
 \mathcal{A}^0 = \mathcal{A}^{0}_{S,-} + \mathcal{A}^{0}_{P,-} + \mathcal{A}^{0}_{S,+} + \mathcal{A}^{}_{P,+} + \mathcal{A}^{0}_{S,0} + \mathcal{A}^{0}_{P,0} 
  +  \cdots  \ ,
\end{equation}
where the ellipsis denotes $D$- and higher-wave contributions and possible high invariant-mass contribution.

\subsection{Amplitudes  for  $\boldsymbol{  D^0 \to K^0_S K^+ K^- }$}
\label{QtbAmpD0K0KK}

The  $D^0 \to K^0_S K^+ K^-$ decay channel was measured by the  BABAR Collaboration with high statistics~\cite{SanchezPRL105_081803}. Using the quasi two-body 
factorization approach~\cite{BLoiseau_Rio2015}, we parametrize~\cite{JPD_inprogress} this decay channel with the definitions of the invariants  $s_{12}$, $s_{13}$ and $s_{23}$ 
similar to  those introduced for  $D^0 \to K^0_S \pi^+ \pi^- $ in the previous section, replacing charged pions by charged kaons, the effective Hamiltonian being identical (here the charged kaon mass is denoted $m_K$). 
The momenta are thus labelled as $D^ 0(p_D)\to K^0_S(p_1) K^-(p_2) K^+(p_3)$ with  $s_{12}=(p_1+p_2)^2$, $s_{13} =(p_1+p_3)^2$,  $s_{23}=(p_2+p_3)^2$, 
and $s_{12}+ s_{13}+s_{23}= m_{D^0}^2+m_{K^0}^2+2 m_K^2$. The involved three interacting kaon pairs, $[K^+ K^-]_L$, $[K^0_S K^-]_L$ and $[K^0_S K^+]_L$ can be 
in a scalar or vector state with $L=S$ or $P$, respectively. The  isospin of the $[K^+ K^-]_L$ pairs can be either 0 or 1, but that of the $[K^0_S K^\mp]_L$ pairs is 1.

\subsubsection{Parametrization of the $D^0 \to K^0_S [K^+ K^-]_S $ amplitudes}
\label{KKS}

The decay amplitude in which the isoscalar $[K^+ K^-]_S$ pair is associated with the $f_0(980)$ and $f_0(1370)$ resonances, and the isovector one is related to 
the $a_0(980)^0$ and $a_0(1450)^0$ resonances, can be parametrized as: 
\begin{equation}
\label{M1kT}
{\mathcal{A}}^0_{S,0}(s_{23}) =  \left({h}_1^S + {h}_2^S s_{23} \right) F^{ K\bar K}_{0n} (s_{23} ) + h_3^S \left(m_{K^0}^2-s_{23} \right) F^{ K\bar K}_{0s} (s_{23} )  +
\left({h}_4^S+{h}_5^S  s_{23} \right) G_0^{K \bar K}(s_{23})  \ .
\end{equation}

The  amplitude with the isovector $ [K^0 K^-]_S$  pairs in an $S$-wave state, which include the $a_0(980)^-$ and $a_0(1450)^-$ resonances, can be parametrized as,
\begin{equation}
\label{M3k}
\mathcal{A}_{S,-}^0(s_{12})=\left(h_6^ S+h_7^ S s_{12}\right) G_0^{K \bar K}(s_{12}),
\end{equation}
and the corresponding amplitude associated with the $a_0(980)^+$ and $a_0(1450)^+$ resonances as,
\begin{equation}
\label{M4k}
\mathcal{A}_{S,+}^0(s_{13})=\left[h_8^S \dfrac{F_0^{D K}(s_{13})}{s_{13}}+h_9^ S(m_{K}^2-s_{13})\right]  G_0^{K\bar K}(s_{13}) \ .
\end{equation}
The scalar-isocalar form factors, $F^ {K\bar K}_{0n(s)}(s)$, in Eq.~(\ref{M1kT}), and the scalar-isovector ones, $G^{K\bar K}_0(s)$, in Eqs.~(\ref{M3k}) and~(\ref{M4k}) are detailed in Appendix~\ref{KK} 
while the scalar $D$ to $K$  transition form factor $F_0^{D K}(s)$ is defined in Appendix~\ref{transitionff}.

\subsubsection{Parametrization of the $D^0 \to K^0_S [K^+ K^-]_P $ amplitudes}
\label{KKP}

We parametrize this decay amplitude, where the isoscalar and isovector $ [K^+ K^-]_P$  pairs contain contributions from the $\omega(782)$, $\omega(1420)$, $\phi(1020)$,  $\rho(770)^0$,  $\rho(1450)^0$, and $\rho(1700)^0$
resonances, by the expression~\cite{JPD_inprogress}:
\begin{equation}
\label{M5k}
\mathcal{A}_{P,0}^0(s_{12}, s_{13}, s_{23})=\left(s_{12} -s_{13}\right) \left(h_1^ P F_{1u}^{K^+  K^-}(s_{23}) +h_{2}^ P F_{1s}^{K^+ K^-}(s_{23})
\right)  \ .
\end{equation}
Likewise, one can express the amplitude in which the isovector  $ [K^0_S K^-]_P$  pair is associated with the three $\rho^-$ resonances by,
\begin{equation}
\label{M8k}
\mathcal{A}_{P,-}^0(s_{12}, s_{13}, s_{23})= h_3^ P \left [s_{23} - s_{13} + \left(m_{D^0}^2- m_{K}^2\right)\ \frac{m_{K^0}^2-m_{K}^2}{s_{12}} \right ]    F_{1}^{K^-  K^0}(s_{12})  \ ,
\end{equation}
while the parametrization of the amplitude associated with the $\rho^+$ resonances  reads,
\begin{equation}
\label{M9k}
\mathcal{A}_{P,+}^0 (s_{12}, s_{13}, s_{23})=  \left[ h_4^ P+h_5^P F_1^{D K}(s_{13}) \right] \left[s_{23} - s_{12} + \left(m_{D^0}^2- m_{K}^2\right)\ \frac{m_{K^0}^2-m_{K}^2}{s_{13}} \right ]  F_{1}^{K^+\bar K^0}(s_{13})\ .
\end{equation}
In Eqs.~(\ref{M5k}) to (\ref{M9k}), the vector-isocalar, $F^ {K\bar K}_{1u(s)}(s)$, the vector-isovector $F_1^{K^-  K^0}(s)$ and  $F_1^{K^+ \bar K^0}(s)$ form factors
are defined in Appendix~\ref{KK}. The parametrization of the 
vector $D$ to $K$ transition form factor $F_1^{D K}(s)$ is discussed in Appendix~\ref{transitionff}.

The full decay amplitude is the coherent sum of all the sub-amplitudes  discussed above:
\beq
 \mathcal{A}^0 = \mathcal{A}^{0}_{S,-} + \mathcal{A}^{0}_{P,-} + \mathcal{A}^{0}_{S,+} + \mathcal{A}^{0}_{P,+} + \mathcal{A}^{0}_{S,0} + \mathcal{A}^{0}_{P,0}  +  \cdots  \ ,
\eeq
where the ellipsis denotes the omission of higher waves that could be included using Breit-Wigner line shapes.


\section{Concluding remarks}
\label{Conclusion}

 We have introduced alternatives to the isobar-model Dalitz-plot
 parametrizations of weak $D$ and $B$ decays into exclusive final
 states composed of three light mesons, namely the various charge
 states $\pi\pi\pi$, $K\pi\pi$ and $KK\bar K$. Such isobar
 parametrizations have been frequently employed in fits although they
 do not respect unitarity, which leads, amongst other effects, to a
 sum of branching fractions that can exceed the total decay width by
 large amounts.  As a consequence, any strong CP phases that may be
 extracted from these fits must be taken with caution.

Our alternative parametrizations, while not fully three-body unitary,
are based on a sound theoretical application of QCD factorization to a
hadronic quasi two-body decay. We thus assume that the final three-meson
state is preceded by intermediate resonant states which is justified
by ample phenomenological and experimental evidence. Analyticity, unitarity, chiral
symmetry as well as the correct asymptotic behavior of the two-meson
scattering amplitude in $S$ and $P$ waves are implemented via analytical and 
unitary $S$- and $P$-wave $\pi\pi$, $\pi K$ and $K\bar K$ form factors which
enter the hadronic final states of our amplitude parametrizations. These
amplitudes can be readily used adjusting the parameters in a least-square
fit to the Dalitz plot --- for a given decay channel --- and employing
tabulated form factors as functions of momentum squared or
energy. The different quasi-two-body $B$- and $D$-decay channels 
for which we provide explicit amplitude expressions are summarized in 
Tables~\ref{ParamBdecay} and~\ref{ParamDdecay}, respectively.
For each channel the relevant equation for the parametrization  
is cited and the dominant contributing resonances are listed.
Let us add a practical remark: in any application of the
parametrized amplitudes to experimental analyses, one can set to
zero one phase of the $S$ or $P$ wave amplitude since the Dalitz plot
density is not sensitive to its value.

With this  ``tool kit" we strongly hope to contribute to more
sophisticated experimental extractions of three-body decay observables, in
particular CP-violating phases.

\begin{table}[t!]
\caption{For  each $B$-decay channel  in the first column, the second column refers to the equation of the  proposed amplitude parametrization and the third column lists the dominant contributing resonances.}
\label{ParamBdecay}
\begin{center}
\begin{tabular}{lcr}
\hline
\hline
Quasi-two-body channel\hspace{0.3cm} & see Eq.:& Dominant resonances \\
\hline
$B^- \to \pi^- [\pi^+\pi^-]_{S}$ & (\ref{MSsij}) & $f_0(500)$, $f_0(980)$, $f_0(1400)$  \\
$B^- \to \pi^- [\pi^+\pi^-]_{P}$ & (\ref{MPsij}) & $\rho(770)^0$, $\rho(1450)^0$ , $\rho(1700)^0$ \\
$B \to K [\pi^+ \pi^-]_{S}$ &(\ref{paracs}) & $f_0(500)$, $f_0(980)$, $f_0(1400)$  \\
$B \to K [\pi^+ \pi^-]_{P}$ & (\ref{paracp}) &  $\rho(770)^0$, $\rho(1450)^0$ , $\rho(1700)^0$ \\
$B^{-(0)} \to [K^{-(0)}\pi^+]_S\pi^-$ & (\ref{MSparam}) & $K^*_0(800)^{0(+)}$, $K^*_0(1430)^{0(+)}$\\
$B^{-(0)} \to [K^{-(0)} \pi^+]_P\pi^-$ & (\ref{MPparam}) & $K^*(892)^{0(+)}$, $K^*(1410)^{0(+)}$ \\
$B^- \to K^- [K^+ K^-]_S$ & (\ref{ASsij}) & $f_0(980)$, $f_0(1400)$\\
$B^- \to K^- [K^+ K^-]_P$ & (\ref{APsij})& $\rho(770)^0$, $\rho(1450)^0$, $\rho(1700)^0$,   \\
    &  &
    $\omega(782)$, $\omega(1420)$, $\omega(1650)$, \\
   &  &  $\phi(1020)$, $\phi(1680)$\\
\hline
\hline
\end{tabular}
\end{center}
\end{table}


\begin{table}[h!]
\caption{As in Table~\ref{ParamBdecay} but for hadronic quasi-two-body $D$ decays.}
\label{ParamDdecay}
\begin{center}
\begin{tabular}{lcr}
\hline
\hline
Quasi-two-body channel \hspace{0.5cm} & see Eq.: & Dominant resonances \\
\hline
$D^+ \to [\pi^+ \pi^-]_S \pi^+ $ & (\ref{AD3piS})& $f_0(500)$, $f_0(980)$, $f_0(1400)$\\
$D^+ \to [\pi^+ \pi^-]_P \pi^+ $ & (\ref{AD3piP}) & $\rho(770)^0$, $\rho(1450)^0$ \\
 $D^+ \to [K^- \pi^+]_S \ \pi^+ $ &(\ref{ASPD+1})&  $K^*_0(800)^{0}$, $K^*_0(1430)^{0}$\\
  $D^+ \to [K^- \pi^+]_P \ \pi^+ $ &(\ref{ASPD+2})& $K^*(892)^0$, $K^*(1410)^0$\\
  $D^0 \to [K^0_S \pi^-]_{S}\ \pi^+ $&(\ref{ASD0}) & $K^*_0(800)^{-}$, $K^*_0(1430)^{-}$\\
   $D^0 \to [K^0_S \pi^-]_{P}\ \pi^+ $ &(\ref{APD0})& $K^*(892)^-$, $K^*(1410)^-$\\
   $D^0 \to [K^0_S \pi^+]_{S}\ \pi^- $ &(\ref{A8SPD0})& $K^*_0(800)^{+}$, $K^*_0(1430)^{+}$\\
    $D^0 \to [K^0_S \pi^+]_{P}\ \pi^- $ & (\ref{A9SPD0})&$K^*(892)^+$, $K^*(1410)^+$ \\
    $D^0 \to K^0_S [\pi^+ \pi^-]_{S}$ &(\ref{A2SPD0}) &  $f_0(500)$, $f_0(980)$, $f_0(1400)$\\
  $D^0 \to K^0_S [\pi^+ \pi^-]_{P}$  & (\ref{A4SPD0})& $\rho(770)^0$, $\omega(782)$\\
  $D^0 \to K^0_S [K^+ K^-]_{S}$  &(\ref{M1kT})&$f_0(980)$, $f_0(1400)$, $a_0(980)^0$,  $a_0(1450)^0$\\
  $ D^0 \to K^+[K^0 K^-]_S$ &(\ref{M3k})& $a_0(980)^-$,  $a_0(1450)^-$\\
  $ D^0 \to K^-[K^0 K^+]_S$ &(\ref{M4k})& $a_0(980)^+$,  $a_0(1450)^+$\\
    $D^0 \to K^0_S [K^+ K^-]_{P}$ &(\ref{M5k})&  $\omega(782)$, $\omega(1420)$, $\phi(1020)$, $\rho(770)^0$, $\rho(1450)^0$  \\
     $ D^0 \to K^+[K^0 K^-]_P$& (\ref{M8k})& $\rho(770)^-$, $\rho(1450)^-$  \\
     $ D^0 \to K^-[K^0 K^+]_P$ &(\ref{M9k})& $\rho(770)^+$, $\rho(1450)^+$\\
\hline
\hline
\end{tabular}
\end{center}
\end{table}


\section*{Acknowledgements}
It is a pleasure to thank the Centro Brasileiro de Pesquisas F\'isicas
(CBPF) and the organizers of the ``2015 LHC$b$ workshop on multi-body
decays of $D$ and $B$ mesons'' where this work was initiated. BL and
DB appreciated the hospitality of Universidade Cruzeiro do Sul where
part of this work was carried out.  Discussions with Alberto dos Reis, Gast\~ao Krein and Patr\'icia Magalh\~aes are gratefully acknowledged.  This work was
partially supported by the S\~ao Paulo Research Foundation (FAPESP), grant
nos. 2015/20689-9 and 2016/03154-7; by the Brazilian National Council for
Scientific and Technological Development (CNPq), grant
nos.~305431/2015-3 and 458371/2014-9; by the IN2P3-COPIN collaboration agreement (project No
08-127); by 
the Spanish ``Ministerio de Econom\'ia y Competitividad" under grants 
CICYT-FEDER-FPA 2014-55613-P and SEV-2012-0234, 
as well as by  Secretaria d'Universitats i Recerca del Departament d'Economia i Coneixement 
de la Generalitat de Catalunya under grant 2014 SGR 1450.


\appendix

\section{Form factors}
\label{AppFFs}

In quantum field theory it can be shown, using dispersion
relations~\cite{Barton65}, that strong interaction meson-meson form
factors can be in principle calculated exactly by means of the
coupled-channel Muskhelishvili-Omn\`es (MO) equations~\cite{MO},
provided one knows the meson-meson scattering matrices at all
energies.  In practice, our knowledge about scattering phases is
incomplete and one has to resort to simplifications. Eventually, different
approaches to the calculation of these form factors lead to 
slightly different results.  In the following, we briefly describe
several state-of-the-art descriptions of the various form factors
employed in the decay-amplitude parametrizations presented in this
work. These form factors can be obtained from the authors of the
original works in the form of numerical tables and be readily employed
in a concrete Dalitz-plot analysis.


\subsection{\boldmath $\pi\pi$ form factors}
\label{Fpp}

The parametrizations of the amplitudes $B^- \to \pi^- [\pi^+\pi^-]_{S,P}$  in Eqs.~(\ref{MSsij}) and (\ref{MPsij}), $B \to K
[\pi^\pm \pi^\mp]_{S,P}$ in Eqs.~(\ref{paracs}) and (\ref{paracp}), $D^+ \to \pi^+[\pi^-\pi^+]_{S,P}$ in Eqs.~(\ref{AD3piS}) and (\ref{AD3piP}),
and $D^0 \to K^0_S [\pi^+ \pi^-]_{S,P}$ in Eqs.~(\ref{A2SPD0}) and (\ref{A4SPD0}) require the knowledge of the pion non-strange scalar 
form factor, $F_{0n}^{\pi\pi} (s_{ij})$, and vector form factor, $F_1^{\pi^+ \pi^-}(s_{ij})$.  The strange pion scalar form factor
$F_{0s}^{\pi\pi} (s_{ij})$ enters the parametrization of the $B \to K [\pi^\pm \pi^\mp]_{S}$ amplitude in Eq.~(\ref{paracs}).

\subsubsection{Scalar form factors}
\label{SFpp}

The scalar form factors $F_{0n(s)}^{\pi \pi}(s_{ij})$ can be found, for example, in Refs.~\cite{fkll,JPD_PRD89, Moussallam_2000,Dedonderetal2011,DHK16}.\footnote{In 
Refs.~\cite{fkll,Dedonderetal2011,ElBennich2006,JPD_PRD89,Boitoetal09} the form factor is defined as  $\Gamma_1^{n(s)*} (s_{ij})=\sqrt{3/2}\ F_{0n}^{\pi \pi}(s_{ij})$ with $F_{0n}^{\pi\pi}(0)=1$. 
The relation for the strange case is ambiguous as $F_{0s}^{\pi\pi}(0)=0$ in the lowest order of chiral symmetry (see Refs.~\cite{meis01,Moussallam_2000} for more details). }
In Ref.~\cite{Dedonderetal2011} the form factors have been derived using a unitary relativistic coupled-channel model including $\pi \pi$, $K
\bar K$ and effective $(2\pi)(2\pi)$ interactions together with chiral symmetry constraints (an approach put forward in Ref.~\cite{meis01}). The latest version of 
the corresponding non-strange form factors was obtained in Ref.~\cite{JPD_PRD89}, with constrains from the high-statistics Dalitz plot data of the $D^0 \to
K^0_S \pi^+ \pi^-$ from Ref.~\cite{BelleDKspipi,SanchezPRL105_081803}.  In this approach the $\pi \pi $ $T$-matrix is  that of the solution $A$ of the three coupled-channel model of
Ref.~\cite{Kaminski_EPJC9_141}, where the effective mass is $m_{(2\pi)}$= 700 MeV.

As an alternative, one can employ the scalar pion form factors obtained from the numerical solution of a coupled channel MO
problem, as derived in Ref.~\cite{Moussallam_2000}. This approach has been recently revisited in the context of $B^0\to J/\psi
\pi \pi$ decays in Ref.~\cite{DHK16}. There, the system of MO equations is solved with input from chiral symmetry constrained by
recent lattice data. These form factors suffer from an uncertainty that stems from the kaon form-factor normalization at zero (which
enters through the coupled-channel equations).  This theoretical uncertainty is more pronounced in the scalar pion form factor at
energies above 800~MeV. 

The modulus of the pion non-strange scalar form factor is characterized by a dip arising from the $f_0(980)$ contribution and by
two bumps whose origin are the $f_0(500)$ and $f_0(1400)$ resonances. The strange scalar form factor is dominated by a peak
around the $f_0(980)$ contribution. The form factors are depicted, for instance, in Fig.~1 of Ref.~\cite{Dedonderetal2011} for the non-strange 
scalar form factor and in Fig.~6 of Ref.~\cite{DHK16} for both, the strange and non-strange scalar case.

\subsubsection{Vector form factor}
\label{VFpp}

The pion vector form factor can be extracted accurately from experimental data for $\tau^- \to \pi^-\pi^0 \nu_\tau$ and $e^+e^-\to
\pi^+\pi^-$.  However, while in the $\tau^-$ decay the current has only  isospin-1 component, the $e^+e^-$ annihilation also implies an isoscalar component.
Recent descriptions can be found, for example, in Refs.~\cite{Fujikawa_PRD78_072006,Hanhart,RoigDumm}.

A good fit to $D^0 \to K^0_S \pi^- \pi^+$ decay data is obtained in Ref.~\cite{JPD_PRD89} using the vector form-factor parameterization
employed by the Belle Collaboration in their data analysis of $\tau^- \to \pi^-\pi^0 \nu_\tau$ decays~\cite{Fujikawa_PRD78_072006}. It is
based on a Gounaris-Sakurai form and the parameters used are those of Table~VII of Ref.~\cite{Fujikawa_PRD78_072006}. The Dalitz plot 
is also well described by the unitary  parametrization of Ref.~\cite{Hanhart}.

Another recent unitary description that can be useful in data analysis is the dispersive representation of Ref.~\cite{RoigDumm}. This
description of the form factor uses Belle data on the $\tau \to \pi\pi \nu$ decays to constrain a three-time subtracted dispersive representation.

Finally, care must be exercised to correctly take into account both the isosvector and isoscalar components. For instance, in $\Dppp$ decays,
the current that couples to the $\pi^+\pi^-$ pair in a $P$-wave is $\bar d \gamma_\mu d$, which contains both isospin 0 and 1. 
One therefore expects the $\omega$ contribution to be sizeable in high-statistics data sets. The inclusion of the $\omega$ contribution
can be done as discussed in detail in Ref.~\cite{DHK16} [see in particular their Eq.~(3.7)]. An alternative  is to take into account  the
contribution of the $\omega$ using the respective isobar model amplitude, described in terms of a Breit-Wigner parametrization.


\subsection{\boldmath $K\pi$ form factors}
\label{Kp}

The $K\pi$ scalar form factor, $F_0^{K\pi}$, and the $K\pi$ vector form factor, $F_1^{K\pi}$,  enter our parametrizations of the 
$B \to [K   \pi^\pm]_{S,P} \,\pi^\mp$, $D^+ \to [K^- \pi^+]_{S,P}\ \pi^+$, and $D^0 \to [K^0_S \pi^\mp]_{S,P}\ \pi^\mp $ amplitudes. 
Below we discuss the determination of these form factors.

\subsubsection{$K\pi$ scalar form factor}
\label{SFKp}

Sophisticated computations of the scalar $F_0^{K\pi}$ form factor by
means of a coupled-channel dispersive representation can be found in
Refs.~\cite{Moussallam:2007qc,JOP}.  The form factor derived in Ref.~\cite{Moussallam:2007qc}
from two coupled-channel MO equations
depends on the ratio $r_{K\pi}=f_K/f_\pi$, $f_K$ and $f_\pi$ being the
kaon and pion decay constants, and was used with success in
Refs.~\cite{ElBennichetal09, JPD_PRD89}. It contains the contributions
of the $K^*_0(800)$ [or $\kappa (800)$]~\cite{SDGBM06} and
$K^*_0(1430)$ resonances clearly visible as bumps. Its modulus
is plotted in Fig.~2 of Ref.~\cite{ElBennichetal09}. 

The same form factor was derived in a coupled-channel ($K\pi$,
$K\eta$, and $K\eta'$) dispersive framework imposing constraints from
Chiral Perturbation Theory at low-energies in Ref.~\cite{JOP}.  The
form factors are obtained from the numerical solution of the
coupled-channel equations with input from the $T$-matrix elements
previously calculated in Ref.~\cite{JOPKpiScattering}. This is the form
factor that was employed in the description of $\DKpp$ decays in
Ref.~\cite{BE2009}.

\subsubsection{$K\pi$ vector form factor}
\label{PFKp}

The $K\pi$ vector form factor can be extracted with accuracy from the
spectrum of $\tau \to K \pi \nu$ decays. These decays are largely
dominated by the vector contribution and the present statistics allows
for a description with good precision. 
The unitary form factor derived in Ref.~\cite{Moussallam:2007qc} from three coupled-channel equations has been used in Ref.~\cite{ElBennichetal09}.
In Refs.~\cite{Boito:2008fq,Boito:2010me}, the
form factor is described by a dispersive relation with three
subtractions and constrained by the Belle data for $\tau^- \to K_S \pi^-
\nu_\tau$~\cite{BelleTauKpinu} and information from $K_{l3}$ decays. The
$K^*(892)$ and $K^*(1410)$ resonances contribute to this vector form
factor. The contribution of the $K^*(1680)$ is difficult to assess due to the larger error bars around 1600 GeV in the spectrum of $\tau^- \to K_S \pi^-
\nu_\tau$. This
form factor has been employed with success in the description of
$\DKpp$ decays of Ref.~\cite{BE2009}.  It also leads to a good fit of
the present high statistics $D^0 \to K^0_S \pi^+ \pi^-$
data~\cite{JPD_PRD89}.

\subsection{\boldmath $KK$ form factors}
\label{KK}

\subsubsection{Scalar-isoscalar case}

The kaon non-strange and strange scalar and isoscalar form factors,
$F_{0n(s)}^{K \bar K}(s_{ij})$,\footnote{In Refs.\cite{fkll,PLB699_102,LZ2014}
these form factors are also defined as $\Gamma_2^{s*} (s_{ij})=  F_{n(s)}^{K K}(s_{ij})/\sqrt{2}$ with $F_{0n}^{KK}(0)=F_{0s}^{KK}(0)=1$ (see Refs.~\cite{meis01,Moussallam_2000} for more details). }  enter the $B^- \to K^-[K^+ K^-]_{S}$ amplitude in
Eq.~(\ref{ASsij}), and the $D^0 \to K^0_S [K^+ K^-]_{S}$ amplitude in
Eq.~(\ref{M1kT}).  They have been calculated
in Ref.~\cite{PLB699_102} with the three coupled channels $\pi\pi,
\bar KK$ and $4\pi$ (effective $(2\pi)(2\pi)$ or $\sigma\sigma$ or
$\eta \eta$, etc.)  in the approach developed in
Ref.~\cite{Dedonderetal2011} to derive the pion scalar form factors (see
Appendix~\ref{SFpp}). Through their coupling to $K \bar K$, the
resonances $f_0(980)$ and $f_0(1400)$ contribute to $F_{0n(s)}^{K \bar
  K}(s_{ij})$, as can be seen from the spikes present in Fig.~1 of
Ref.~\cite{PLB699_102}.  An alternative derivation of these
form factors using MO equations has been presented in
Ref.~\cite{Moussallam_2000} and represents a sound alternative.

\subsubsection{Scalar-isovector case}

For an isospin 1 $[K^+ K^-]$ pair and assuming isospin symmetry, the scalar-isovector form factor $G_0^{K \bar K}(s)=G_0^{[K^+ K^-]}(s) =G_0^{K^0 K^-}(s) =G_0^{\bar K^0 K^+}(s)$ is defined as~\cite{EPJC_75_488},
 \begin{equation}
\label{G0KK}
\mathcal{B}^0 G^{K \bar K}_0(s)=\langle \bar K^0(p_{K^-})K^+(p_{K^+})|\bar u d |0 \rangle \ ,
\end{equation}
with $\mathcal{B}^0=m_\pi^2/(m_u+m_d)$.  This form factor, entering the $D^0 \to K^0_S [K^+ K^-]_{S}$ amplitude in
Eq.~(\ref{M1kT}), was calculated in Ref.~\cite{EPJC_75_488} from coupled MO equations for $\pi \eta$ and $K \bar K$ channels.  
The above form factor includes the contributions of the $a_0(980)$ and $a_0(1450)$ seen as bumps in their moduli (see for instance 
the right panel of Fig.~7 of Ref.~\cite{EPJC_75_488}).

\subsubsection{Vector case}
For the $B^- \to K^- [K^+ K^-]_P$ amplitude in Eq.~(\ref{APsij}) and  for the $D^0 \to K^0_S [K^+ K^-]_{P}$ amplitude in Eq.~(\ref{M5k}), the vector form factors $F^{K^+K^-}_{1q}(s)$ with $q=u,d$ and $s$ 
are defined through~\cite{Bruch}:
\begin{equation}
  \label{Fq}
  \langle K^+(p_i)K^-(p_j)|\bar q \gamma_{\nu} q |0 \rangle =  (p_i-p_j)_{\nu}\, F^{K^+K^-}_{1q}(s_{ij}) \ .
\end{equation}
They have been calculated using vector dominance, quark model
assumptions and isospin symmetry in Ref.~\cite{Bruch} and receive
contributions from the eight vector mesons: $\rho(770)$, $\rho(1450)$,
$\rho(1700)$, $\omega(782)$, $\omega(1420)$, $\omega(1650)$,
$\phi(1020)$ and $\phi(1680)$. The form factor can be written in closed form using, for example, Eqs. (23) to (25) of Ref.~\cite{PLB699_102}. The parameters needed can be obtained from Table~2 of Ref.~\cite{Bruch}.

The isovector $K\bar K$ form factors that enter the amplitudes   $D^0 \to K^\mp [K^0_S K^\pm]_{P}$ are defined
\begin{eqnarray}
\label{FUK} 
 \langle K^+(p_i) \bar K^0(p_j)| \bar u \gamma_\nu d|0\rangle &=&(p_i-p_j)_{\nu} \, F_1^{K^+ \bar K^0}(s_{ij}), \\
\langle K^-(p_i) K^0(p_j)| \bar d \gamma_\nu u|0 \rangle&=&(p_i-p_j)_{\nu}\, F_1^{K^- K^0}(s_{ij}).
\end{eqnarray}
Using isospin symmetry one can obtain the following relations~\cite{Bruch}
\beq
 F_1^{K^+ \bar K^0}(s_{ij})= - F_1^{K^- K^0}(s_{ij})= 2 F_{1u,I=1}^{K^+K^-}(s_{ij}),
\eeq
where  $F_{1u,I=1}^{K^+K^-}(s_{ij})$ is the $I=1$ component of the charged kaon form factor. This form factor is described by Eq.~(23) of Ref.~\cite{PLB699_102} keeping only the $\rho$ meson contributions.

\subsection{Heavy-to-light transition form factors}
\label{transitionff}

As discussed in the introduction, factorization theorems allow to perturbatively integrate out energy scales and yield approximations which are exact
in the infinite heavy-quark limit. To a reasonable extent, the decay amplitudes factorize in terms of products of hard and soft matrix elements. Amongst the latter,
heavy-to-light transitions factors have been extensively studied in the past two decades, though their precise nonperturbative evaluation remains a challenge. 
Full {\em ab initio\/} calculations valid in any momentum-squared region are currently out of reach and one is mostly left with modelling the heavy-to-light amplitudes 
with as much input from nonperturbative QCD as possible; in many cases, form factors are only obtainable for a limited range of momentum squared, $q^2$, 
values and then extrapolated to other $q^2$ values.

The transition amplitude of a heavy pseudoscalar meson $H$ to a lighter pseudoscalar meson $P$ via an electroweak current, $\langle P (p_P) | J_{\mu} | H (p_H) \rangle $, 
is described  by two dimensionless form factors,
\begin{equation}
   \langle P (p_P) | \bar l\, \gamma_\mu (1- \gamma_5) h | H(p_H)  \rangle = F_+ (q^2) (p_H + p_P)_\mu + F_-(q^2) (p_H- p_P)_\mu  \ ,
\end{equation}
where $l = u,d,s$, $h=c,b$ and where the transferred momentum is $q=p_H-p_P$. It is convenient to rewrite this amplitude in terms of  another pair of form factors, 
namely the scalar and vector form factors, $F_0(q^2)$ and $F_1(q^2)$, respectively introducing the momentum $K = p_H+p_P$~\cite{Lu:2007sg,Melikhov:2001zv,Ivanov:2007cw}:
\begin{equation}
\langle P  (p_P)\vert \bar l\, \gamma_\mu (1-\gamma_5) h \vert H (p_H)  \rangle  
               = F_1 (q^2)   \left [ K_\mu -\frac{K\cdot q }{q^2} \, q_\mu  \right ]  +  F_0 (q^2)\, \frac{K\cdot q }{q^2} \, q_\mu \ .
 \label{scalvec}
\end{equation}
The relation between the two sets of form factors is given by,
\begin{align}
  F_1(q^2) & = F_+(q^2)  \ ,  \\
  F_0(q^2) & = F_+(q^2)  + F_-(q^2)\, \frac{q^2}{K\cdot q } \ , 
\end{align}
where  at $q^2 = 0$ the identity, $F_1(0) = F_0(0) = F_+(0)$, holds. Notice that the above definitions are identical for the $\langle S (p_S) | J_{\mu} | H (p_H) \rangle $
transitions, e.g. when the final state $S$ is a scalar meson.

The advantage of the Lorentz decomposition in Eq.~(\ref{scalvec}) lies in the simplification of the decay amplitudes:  if the meson, emitted via an electroweak gauge boson,
is a pseudoscalar (or scalar), then only $F_0(q^2)$ enters the decay amplitude. Analogously, if the emitted meson is a vector (or axial-vector) meson,
the decay amplitude only depends on $F_1(q^2)$.

The weak transition of a heavy pseudoscalar meson $H$ to a lighter vector meson $V$ can be decomposed into Lorentz invariants as~\cite{Ali1998},
\begin{align}
\label{heavyvector}
\langle V(p_V,\epsilon_V) \, |  \,\bar l \, \gamma_\mu(1-\gamma_5)\, b\, | H (p_H )   \rangle  =  
                  \frac{-2V}{m_H+m_V }\,  \varepsilon_{\mu\nu\alpha\beta} \, & \epsilon_V^{*\nu}\, p_H^\alpha p_V^\beta  
                  - 2i m_V A_0(q^2) \frac{\epsilon_V^* \cdot q }{q^2} \,q_\mu  \\
                  - i(m_H +m_V ) \, A_1(q^2) \left [ \epsilon_{V\mu}^*   -   \frac{\epsilon_V^* \cdot q }{q^2} \, q_\mu  \right ]  
                 + i A_2(q^2) & \frac{\epsilon_V^* \cdot q }{m_H +m_V }  \left [ ( p_H + p_V )_\mu - \frac{m_H^2 - m_V^2}{q^2}\, q_\mu \right ] \ , \nonumber
\end{align}
where $\epsilon_V$ is the polarization of the final-state vector meson, $q=p_H-p_V$, $p_V^2 = m_V^2$ and $p_H^2=m_H^2$. Other, related decompositions are possible; 
see, e.g.,  Ref.~\cite{Lu:2007sg,Ivanov:2007cw,Melikhov:2001zv,Khodjamirian:2006st,Ball:2004rg,Leitner:2010fq,Paracha:2014wra}. Their relations with the form factor decomposition in 
Eq.~(\ref{heavyvector}) is detailed in Ref.~\cite{Ivanov:2007cw} where algebraic interpolations for the transition form factors can also be found.

A variety of theoretical approaches have been applied to the transition form factors in Eqs.~(\ref{scalvec}) and (\ref{heavyvector}), amongst which are 
analyses using light-front and relativistic constituent quark models, light-cone sum rules, continuum functional QCD approaches and lattice-QCD simulations. 
An experimental extraction of the transition form factors from semi-leptonic decays for a range of $q^2$ momenta is possible and has been obtained, for instance, 
in the case of $D^0 \to \pi^- e^+ \nu_e$ decays~\cite{Lees:2014ihu}.
These decays are considerably easier to analyze than non-leptonic decays final states entail complicated final-state interactions. For a brief summary 
of the theoretical approaches we refer to Ref.~\cite{ElBennich:2009vx}, where a numerical comparison of the theoretical transition form factor, $F_+^{B\to\pi}(q^2)$,
predictions for various $q^2$ values is provided in Table~1 and which highlights pronounced variations amongst the approaches. A comparison of numerical  
results for the $B\to K^*$ form factors obtained in lattice-QCD, light-cone sum rules and Dyson-Schwinger equation approaches is presented in Fig.~2 of 
Ref.~\cite{Paracha:2014wra}.


\section{Relations between the parametrized and original amplitudes}
\label{relationfreeoriginal}

The aim of this appendix is to relate the amplitudes introduced in
Sections~\ref{3bodyBdecayAmp} and \ref{3bodyDdecayAmp} to those
derived in quasi-two-body QCD factorizations
~\cite{Dedonderetal2011,fkll,ElBennich2006,ElBennichetal09,PLB699_102,Boitoetal09,BE2009,JPD_PRD89}
which represents the original motivation of the present
parametrizations.  The main purpose of this section is to make contact
with the original works and make explicit the physical meaning behind
the different parameters of the amplitudes we discussed here.  The
relations are presented following the order of appearance of the
three-body decay amplitudes in Sections~\ref{3bodyBdecayAmp} and
\ref{3bodyDdecayAmp}. Explanations of and details about constants and
form factors that occur in the amplitudes below can be found in the
original references we quote.

\subsection{$\boldsymbol B$-decay amplitudes}
\label{Bda}

In the parameters below, when necessary, the superscripts $-, \bar 0, +$ and $0$ refer to the $B^-, \bar B^0, B^+$  and $B^0$ mesons, respectively.   
 
\subsubsection{$B^\pm \to\pi^\pm  \pi^+ \pi^- $}
\label{relaBppp}

Comparing the parametrized $B^- \to \pi^- [\pi^+ \pi^-]_{S,P} $ $S$ and $P$ amplitudes, Eqs.~(\ref{MSsij}) and (\ref{MPsij}) to the corresponding amplitudes, Eqs.~(22) and (23) 
in Ref.~\cite{Dedonderetal2011} yields,
\begin{eqnarray}
\label{a1S}
a_{1}^S&=&-\frac{G_F}{\sqrt{2}}\chi_S f_\pi F_0^{BR_S}(m_\pi^2) u(R_S \pi^-) \ ,\\
\label{a2S}
a_{2}^S&=& \frac{G_F}{\sqrt{2}} B_0 \frac {M_B^2-m_\pi^2}{m_b-m_d} v(\pi^-R_S) \ ,\\
\label{a1P}
a_{1}^P&=&\frac{G_F }{\sqrt{2}} \ 
N_P\frac{f_\pi}{f_{R_P}} A_0^{B R_P}(m_\pi^2) u(R_P \pi^-)  \, ,\\
\label{a2P}
a_{2}^P&=&\frac{G_F }{\sqrt{2}}\ w(\pi^- R_P) \ .
\end{eqnarray}
 The definitions and numerical values of  all the quantities in Eqs.~\eqref{a1S} to \eqref{a2P} can be found in Ref.~\cite{Dedonderetal2011}.
The functions  $u(R_P \pi^-)$,  $v(\pi^-R_S)$ and $w(\pi^- R_P)$, corresponding  to the short distance contributions, are proportional to the  CKM
matrix elements and to the effective Wilson coefficients. The dominant meson resonances are $R_S\equiv f_0(980)$ and $R_P\equiv \rho(770)^0$ (see Ref.~\cite{Dedonderetal2011}).
Applying CP conjugation to the right-hand side of Eqs.~\eqref{a1S} to \eqref{a2P} yields the relations between the $a_i$ coefficients of the parametrized $B^+ \to \pi^+ [\pi^- \pi^+]_{S,P} $ 
amplitudes to the original amplitude parameters.

\subsubsection{$B \to K \pi^+ \pi^-$}
\label{relaBKpp}

Comparison of the $B^-\to K^- [\pi^+ \pi^-]_{S}$ amplitude given in Eq.~(1) of Ref.~\cite{fkll} with the parametrized form~(\ref{paracs}) leads to 
\begin{eqnarray}
\label{b1S}
b_{1}^{-S}&=& \frac{G_F}{\sqrt{2}}\left[\chi f_K F_0^{B\to(\pi\pi)_S}(m_K^2)\ U
-\tilde C\right],\\
\label{b2S}
b_{2}^{-S}&=&\frac{G_F}{\sqrt{2}} \frac{2\sqrt{2}B_0}{m_b\!-\!m_s}(M_B^2\!-\!m_K^2)V,\\
\label{b3S}
b_{3}^{-S}&=& -\frac{G_F}{\sqrt{2}}\chi \left(M_B^2-m_K^2\right) \ \tilde C,
\end{eqnarray}
where $\tilde C=f_\pi F_\pi\left(\lambda_uP_1^{GIM}\!\!+\!\lambda_tP_1\right)$ with $\lambda_u= V_{ub} V_{us}^*$ and $\lambda_t= V_{tb} V_{ts}^*$.
Furthermore, for $i=1, 2, 3$
\begin{eqnarray}
\label{bib0}
b_{i}^{\bar 0 S}&=& \frac{b_i^{-S}}{\sqrt{2}}, \\
\label{bi+}
b_{i}^{+(0)S}&=& b_i^{-(\bar 0)S}(\lambda_u^*,\lambda_t^*).
\end{eqnarray}
The quantities entering Eqs.~(\ref{b1S}) to (\ref{b3S}) are defined in Ref.~\cite{fkll} where  their numerical values are also given.

The parameter $b_{1}^{-P}$ of the $B^-\to K^- [\pi^+ \pi^-]_{P}$ amplitude~(\ref{paracp}) is related to the parameters described in Ref.~\cite{ElBennich2006} in the following way,
\begin{eqnarray}
\label{b1pmP}
b_1^{\pm P}&=& \frac {A^\pm}{\sqrt{2} m_\rho f_\rho},\\
\label{b10P}
b_1^{0(\bar 0) P}&=& \frac {A^0( {A}^{\bar 0})}{\sqrt{2} m_\rho f_\rho},
\end{eqnarray}
with
\begin{eqnarray}
\label{A-}
A^- &= &G_F\, m_{\rho} [\, f_K\,A_0^{B\to \rho}(M_K^2)\,( U^- - C^P )
  +  f_{\rho}\,F_1^{B\to K}(m_{\rho}^2)\,W^- ],\\
\label{A0}
A^{\bar 0} &=& G_F\, m_{\rho} \left [ f_K\,A_0^{B\to \rho}(M_K^2)\,( U^{\bar 0}+C^P) 
+  
f_{\rho}\,F_1^{B\to K}(m_{\rho}^2)\,  W^{\bar 0} \right ],\\
\label{A+}
A^+(A^0) &=& -A^-( A^{\bar 0}) (\lambda_u^*,\lambda_t^*).
\end{eqnarray}
Definitions and values of the parameters appearing in Eqs.~(\ref{A-}) to (\ref{A+}) can be found in Ref.~\cite{ElBennich2006}.

In Eqs.~(\ref{b1S}), (\ref{b2S}), (\ref{A-}) and (\ref{A0}) the short distance contribution fuctions $U$, $V$, $U^{-(\bar 0)}$ and $W^{-(\bar 0)}$ are product of  CKM quark-mixing matrix elements with effective Wilson coefficients.

Comparing the $B^- \to [K^- \pi^+]_S \pi^-$ and $\bar B^0 \to [\bar K^0 \pi^-]_S \pi^+$ amplitudes given by  Eqs.~(10) and~(14) of Ref.~\cite{ElBennichetal09} with their parametrized forms~(\ref{MSparam}) leads to
\begin{eqnarray}
\label{c1S}
c_1^{-S}&=&\dfrac{G_F}{\sqrt{2}} (M_B^2-m_{\pi}^2)(m_{K}^2-m_{\pi}^2) \nonumber \\
&\times&\left[\lambda_u\left(a_4^{u}(S)-\dfrac{a_{10}^{u}(S)}{2}+c_4^{u}\right)
+\lambda_c\left(a_4^c(S)-\dfrac{a_{10}^c(S)}{2}+c_4^{c}\right)\right],\\
\label{b1m}
c_2^{-S}&=& -\sqrt{2} G_F \dfrac{(M_B^2-m_{\pi}^2)(m_{K}^2-m_{\pi}^2)}{(m_b-m_d)(m_s-m_d)}\nonumber \\
&\times& \left[\lambda_u\left(a_6^{u}(S)-\dfrac{a_8^{u}(S)}{2}+c_6^{u}\right)
+\lambda_c\left(a_6^c(S)-\dfrac{a_8^c(S)}{2}+c_6^{c}\right)\right],\\
\label{b0b0}
c_1^{\bar 0 S}&=&\dfrac{G_F}{\sqrt{2}} (M_{\bar B^0}^2-m_{\pi}^2)(m_{\bar K^0}^2-m_{\pi}^2) \nonumber \\
&\times&\left[\lambda_u\left(a_1+a_4^u(S)+a_{10}^u(S)+c_4^u\right)
+\lambda_c\left(a_4^c(S)+a_{10}^c(S)+c_4^{c}\right)\right],\\
\label{b1b0}
c_2^{\bar 0 S}&=& -\sqrt{2} G_F \dfrac{(M_{\bar B^0}^2-m_{\pi}^2)(m_{\bar K^0}^2-m_{\pi}^2)}{(m_b-m_d)(m_s-m_d)}\nonumber
 \\
&\times& \left[\lambda_u\left(a_6^u(S)+a_8^u(S)+c_6^u\right)
+\lambda_c \left(a_6^c(S)+a_8^c(S)+c_6^c\right)\right],\\
\label{bp0}
c_{1,2}^{+(0) S}&=&c_{1,2}^{-(\bar 0) S}\left(\lambda_u\to\lambda_u^*,\ \lambda_c\to\lambda_c^*\right),
\end{eqnarray}
where  $\lambda_c= V_{cb} V_{cs}^*$ . 

Comparison of the parametrized $K\pi$ $P$-wave amplitude~(\ref{MPparam}) to the original one in  Eqs.~(11) and~(15) of Ref.~\cite{ElBennichetal09} gives  
\begin{eqnarray}
\label{c1m}
c_1^{-P} &=& \dfrac{G_F}{\sqrt{2}}\bigg\{
 \lambda_u\left(a_4^{u}(P)-\dfrac{a_{10}^{u}(P)}{2}+c_4^{u}\right)+ 
 \lambda_c\left(a_4^c(P)-\dfrac{a_{10}^c(P)}{2}+c_4^{c}\right) \nonumber \\
&+&2\dfrac{m_{K^*}}{m_b}\dfrac{f^\perp_V(\mu)}{f_V}\left[ 
 \lambda_u\left(a_6^u(P) -\dfrac{a_8^u(P)}{2} +c_6^u\right)
+\lambda_c\left(a_6^c(P) -\dfrac{a_8^c(P)}{2} +c_6^c\right) \right]
\bigg\},\\
\label{c1b0}
c_1^{\bar 0 P} &=& \dfrac{G_F}{\sqrt{2}}\bigg\{
 \lambda_u\left(a_1+ a_4^u(P)+a_{10}^u(P)+c_4^u\right)+ 
 \lambda_c\left(     a_4^c(P)+a_{10}^c(P)+c_4^c\right)\nonumber\\
&+&2\dfrac{m_{K^*}}{m_b}\dfrac{f^\perp_V}{f_V}\left[ 
 \lambda_u\left(     a_6^u(P) +a_8^u(P)+c_6^u\right)
+\lambda_c\left(     a_6^c(P) +a_8^c(P)+c_6^c\right) \right]
\bigg\}, \\
\label{c1p}
c_1^{+(0) P}&=& c_1^{-(\bar 0) P}\left(\lambda_u\to\lambda_u^*,\ \lambda_c\to\lambda_c^*\right).
\end{eqnarray}
The values and the definitions of the different short range parameters entering Eqs.~(\ref{c1S}) to (\ref{c1b0}) can be found in Ref.~\cite{ElBennichetal09}.
Let us just mention that 
the $a_1$, $a_i^{u(c)}(S/P)$, $i=4, 6, 8, 10$ are leading order factorization (effective Wilson) coefficients to which vertex and penguin corrections are added. The $c_i^{u(c)}$, $i=4, 6$ are free fitted parameters representing non-perturbative and higher order contributions to the penguin diagrams~\cite{ElBennichetal09}.

\subsubsection{$B^\pm \to K^+ K^- K^\pm$}
\label{relaBKKK}

Comparison of the original $B^- \to K^-[K^+ K^-]_{S,P}$ amplitudes (see Eqs.~(2) and (3) of Ref.~\cite{PLB699_102}) with the parametrized forms of Eqs.~(\ref{ASsij}) and (\ref{APsij}) leads to
\begin{eqnarray}
\label{d1S}
d_1^{-S}&=&-\dfrac{G_F}{\sqrt{2}} \chi f_{K}F^{B\to[K^+K^-]_S}_0(m^2_{K})y,\\
\label{d2S}
d_2^{-S}&=&\dfrac{2B_0 G_F}{m_b-m_s} (M^2_B-m^2_{K})v,\\
\label{d1P}
d_1^{-P}&=&\dfrac{G_F}{\sqrt{2}}\dfrac{f_K}{f_{\rho}}A^{B\rho}_0(m^2_{K})y,\\
\label{d2P}
d_2^{-P}&=&-\dfrac{G_F}{\sqrt{2}}w_u,\\
\label{d3P}
d_3^{-P}&=&-\dfrac{G_F}{\sqrt{2}}w_d,\\
\label{d4P}
d_4^{-P}&=&-\dfrac{G_F}{\sqrt{2}}w_s.
\end{eqnarray}
The definition and numerical values of the different parameters entering Eqs.~(\ref{d1S}) to~(\ref{d4P}) can be found in Ref.~\cite{PLB699_102}.
The parameters $y$, $v$, $w_u$, $w_d$ and $w_s$ represent the contribution of the short range weak decay amplitudes.
For the $B^+ \to K^+[K^+ K^-]_{S,P}$ amplitudes one has

\begin{equation}
d_i^{+ S(P)}= d_i^{- S(P)}\left(\lambda_u\to\lambda_u^*,\ \lambda_c\to\lambda_c^*\right).
\end{equation}

\subsection{\boldmath $D$-decay amplitudes}
\label{Dda}

\subsubsection{$D^+ \to \pi^+ \pi^- \pi^+ $}
\label{relaDppp}

The parameters of the $\Dppp$ amplitudes given in Eq.~(\ref{AD3piS}) can be related to the underlying description  of~\cite{Boitoetal09} as follows
\bea
 e_1^{S} &=& \frac{G_F}{\sqrt{2}}V_{cd}V_{ud}^\ast \, a_1 f_\pi \chi_{S}^{\rm eff}, \\
 e_1^{P}  &=& \frac{G_F}{\sqrt{2}}V_{cd}V_{ud}^\ast\,  a_1 f_\pi \chi_{P}^{\rm eff},\\
e_2^{P}  &=& \frac{G_F}{\sqrt{2}}V_{cd}V_{ud}^\ast \, a_2. 
\eea
 The parameters $\chi_{S,P}^{\rm eff}$ are  related
to the contribution of intermediate resonances in the matrix element of the
$a_1$ type~\cite{Boitoetal09}. We
use $f_\pi =\sqrt{2}F_\pi\simeq 130.5~\rm{MeV}$.

\subsubsection{$D^+ \to K^- \pi^+ \ \pi^+ $}
\label{relaDKpp}

The complex parameters of the $\DKpp$ amplitude  given in Eq.~(\ref{ASPD+1}) can be related to the description of Ref.~\cite{BE2009} as
\bea
f_1^{S}&=&\frac{G_F}{\sqrt{2}}V_{cs}V_{ud}^\ast f_\pi \chi_{S}^{\rm eff} a_1,\\
f_1^P &=& \frac{G_F}{\sqrt{2}}V_{cs}V_{ud}^\ast f_\pi \chi_{V}^{\rm eff} a_1,\\
f_2^{S} &=& \frac{G_F}{\sqrt{2}}V_{cs}V_{ud}^\ast \Delta_+^2 a_2,\\
f_2^{P} &=& \frac{G_F}{\sqrt{2}}V_{cs}V_{ud}^\ast \Delta_+^2 a_2.
\eea
The notation and definitions are analogous to the $\Dppp$ case.  We
use again $\Delta^2_+=(m_{K^-}^2 -m_\pi^2)(m^2_{D^+}-m^2_\pi)$.     The parameters $\chi_{V,S}^{\rm eff}$ are related
to the contribution of intermediate resonances in the $a_1$-type
amplitude. We refer to Ref.~\cite{BE2009} for their precise
definition.

As a final comment, experiments found an off-set of about $-65^\circ$
between the $S$- and $P$-wave
phases~\cite{E791DKpipi,FOCUSDKpipi,CLEODKpipi} that is crucial to
reproduce the Dalitz plot~\cite{BE2009}. This off-set in the phases is
described, in the parametrization proposed here,  by  the phases of the  $f_{1,2}^L$ parameters. We
should point out, however, that the dynamical origin of the phase
difference between the $S$- and $P$-wave may be related to hadronic
three-body rescattering that is beyond our description~\cite{Magetal2011} although some controversy persists (see Ref.~\cite{Nakamura}).

\subsubsection{$D^0 \to K^0_S \pi^+\ \pi^- $}
\label{relaD0Kpp}

Comparison between the different $\mathcal{A}_{S(P)}$ amplitudes, Eqs.~(\ref{ASD0}) to (\ref{A4SPD0}) and the   $\mathcal{M}_i$ amplitudes, Eqs.(66-69), (71), (84) and (85)  of Ref.~\cite{JPD_PRD89} yields
the following  relations.

For the $D^0 \to [K^0_S \pi^-]_{S,P}\ \pi^+ $ amplitudes one has
\begin{eqnarray}
\label{g1S}
g_{1}^S&=&\alpha_1 \ m_{D^0}^2 +\beta_1\ m_\pi^2,\\
\label{g2S}
g_{2}^S&=& -(\alpha_1+\beta_1),\\
\label{alpha1}
\alpha_1&=&-\frac{G_F}{2} \ a_1 \ \Lambda_1 \ \chi_1 
 f_\pi \ F_0^{D^0R_S[\bar{K}^0\pi^-]}(m_\pi^2),\\
\label{beta1}
\beta_1&=&-\frac{G_F}{2}\ a_2\ \Lambda_1 \ \chi_1\ f_{D^0} 
\ F_0^{R_S[\bar{K}^0\pi^-] \pi^+}(m_{D^0}^2),\\
\label{g1P}
g_{1}^P&=&\frac{G_F}{2}\ \Lambda_1 \  \left [ a_1 \ \frac{f_\pi}{f_{{K^{*-}}}} \ A_0^{D^0R_P[\bar{K}^0\pi^-]}(m_\pi^2 ) \  - \ a_2  \ \frac{f_{D^0}}{f_{K^{*-}}} \ A_0^{\pi^+ R_P[\bar{K}^0\pi^-] }(m_{D^0}^2) \right ].
\end{eqnarray}

The relations for the $D^0 \to [K^0_S \pi^+]_{S,P}\ \pi^- $ amplitudes are
\begin{eqnarray}
\label{g3S}
g_{3}^S&=&-\frac{G_F}{2} \ \Lambda_2 \ z_8 \ a_2  \ \chi_1 \ f_{D^0} \ \ F_0^{\pi^- R_S[K^0\pi^+]}(m_{D^0}^2),\\
\label{g4S}
g_{4}^S&=&\frac{G_F}{2} \ \Lambda_2 \ z_8\ a_1,\\
\label{g2P}
g_{2}^P&=&- \frac{G_F}{2} \ \Lambda_2 \ z_9\ a_2 \
 \frac{f_{D^0}}{f_{K^{*+}}} \ A_0^{R_P[{K^0}\pi^+] \pi^- }(m_{D^0}^2),\\
\label{g3P}
g_{3}^P&=&- \frac{G_F}{2} \ \Lambda_2 \ z_9\ a_1.
\end{eqnarray}

And for the $D^0 \to K^0_S [\pi^+ \pi^-]_{S,P}$ amplitudes, it reads
\begin{eqnarray}
\label{g5S}
g_{5}^S&=&\alpha_2\ m_{D^0}^2 +\beta_2\ m_{K^0}^2,\\
\label{g6S}
g_{6}^S&=& -(\alpha_2+\beta_2),\\
\label{alpha2}
\alpha_2&=&-  \frac{G_F}{2} \  a_2 \  (\Lambda_1 + \Lambda_2 ) \ \chi_2 
\  f_{K^0} \ F_0^{D^0R_S[\pi^+\pi^-]}(m_{K^0}^2),\\
\label{beta2}
\beta_2&=& -  \frac{G_F}{2} \  a_2 \  (\Lambda_1 + \Lambda_2 ) \ \chi_2 \ 
 f_{D^0} \ F_0^{\bar{K}^0R_S[\pi^+\pi^-]} (m_{D^0}^2),\\
\label{g4P}
g_{4}^P&=&\frac{G_F}{2} \  a_2  \ (\Lambda_1 + \Lambda_2) \  \frac{1}{f_{{\rho}}}
\left [f_{K^0} \ A_0^{D^0R_P[{\pi^+\pi^-}]}(m_{K^0}^2) \  + f_{D^0} \ 
A_0^{\bar{K}^0 R_P[{\pi^+\pi^-}]} (m_{D^0}^2)\right],\\
\label{g5P}
g_{5}^P&=&\frac{G_F}{2} \ (\Lambda_1 + \Lambda_2)   \  \frac{a_2 }{\sqrt{2}} \ 
\left [ f_{K^0} \ A_0^{D^0 \omega}(m_{K^0}^2) \ - \ f_{D^0} \ A_0^{\overline{K}^0  [\pi^+\pi^-]_{\omega}}(m_{D^0}^2) \right ] \frac{g_{\omega\pi\pi}}{m_{\omega}}.
\end{eqnarray}

For the definitions and numerical values of all parameters entering Eqs.~(\ref{g1S}) to (\ref{g5P}) see Ref.~\cite{JPD_PRD89}.

\subsubsection{$D^0 \to K^0_S K^+ K^- $}
\label{relaD0KKK}
Comparison between the parametrized $\mathcal{A}_{S(P)}$ amplitudes, Eqs.~(\ref{M1kT}) to (\ref{M9k}) and the corresponding amplitudes of Ref.~\cite{JPD_inprogress} yields for the kaon pairs in scalar states
\begin{eqnarray}
\label{h1ST}
{h}_{1} ^S&=& - \frac{G_F}{4}\ (\Lambda_1+\Lambda_2) \ a_2\ \chi^{n}\left[f_{K^0}\ 
m_{D^0}^2 F_0^{D^0 f_{0}}(m_{K^0}^2) + \ f_{D^0} \ m_{K^0}^2  F_0^{K^0f_0}(m_{D^
0}^2) \right],\\
\label{h2ST}
{h}_{2} ^S &=& + \frac{G_F}{4}\ (\Lambda_1+\Lambda_2) \ a_2\ \chi^{n}\left[f_{K^0}\
 F_0^{D^0 f_{0}}(m_{K^0}^2) + \ f_{D^0} F_0^{K^0f_0}(m_{D^0}^2) \right],\\
\label{h2S}
{h}_{3} ^S &=& - \frac{G_F}{2}\ (\Lambda_1+\Lambda_2) \ a_2\ \chi^{s} f_{D^0}\
F_0^{K^0 f_{0}}(m_{D^0}^2),\\
\label{h3ST}
{h}_{4} ^S &=& - \frac{G_F}{4}\ (\Lambda_1+\Lambda_2) \ a_2\ \chi^{(1)}f_{K^0}\ m_{D^0}^2 F_0^{D^0 a_{0}^0}(m_{K^0}^2) 
,\\
\label{h4ST}
{h}_{5} ^S&=& + \frac{G_F}{4}\ (\Lambda_1+\Lambda_2) \ a_2\ \chi^{(1)}
f_{K^0}\ 
 F_0^{D^0 a_{0}^0}(m_{K^0}^2) 
,\\
\label{h4S}
{h}_{6} ^S &=&   - \frac{G_F}{2}\ \Lambda_2 \  \left [ a_1 \  f_{K^+}\ m_{D^0}^2 \ F_0^{D^0 a_0^-}(m_{K}^2) +  a_2 \ f_{D^0} \ m_{K}^2  F_0^{K^+a_0^-}(m_{D^0}^2)  \right ],\\
\label{h5S}
{h}_{7} ^S &=&    \frac{G_F}{2}\ \Lambda_2 \  \left [ a_1 \  f_{K^+} \ F_0^{D^0 a_0^-}(m_{K}^2) +  a_2 \ f_{D^0} \  F_0^{K^+a_0^-}(m_{D^0}^2)  \right ],\\
\label{h6S}
{h}_{8} ^S&=&  - \frac{G_F}{2}\ \Lambda_1 \  a_1 \ (m_{D^0}^2- m_{K}^2)\ (m_{K}^2-m_{K^0}^2),\\
\label{fo41}
{h}_{9} ^S&=&  - \frac{G_F}{2} \ \Lambda_1 \ a_2 \ f_{D^0} \  F_0^{K^-a_0^+}(m_{D^0}^2)\ \chi^{(1)}.
\end{eqnarray}

For the kaon pairs in vector states one has
 \begin{eqnarray}
\label{h1P}
h_1^P&=& \frac{G_F}{2}\ (\Lambda_1+\Lambda_2) \ a_2\   \frac{f_{K^0}}{f_{\omega}}\ A_0^{D^0 \omega}(m_{K^0}^2), \\
\label{h2P}
h_2^P&=& \frac{G_F}{2}\ (\Lambda_1+\Lambda_2) \ a_2\   \frac{f_{K^0}}{f_{\phi}}\  A_0^{K^0 \phi}(m_{D^0}^2)), \\
\label{h4P}
h_3^P &=& \frac{G_F}{2}\ \Lambda_2 \ \left [\ a_1\frac{f_{K^+}}{f_{\rho}} A_0^{D^0 \rho^-}(m_{K}^2) - a_2 \frac{f_{D^0}}{f_{\rho}} \ A_0^{K^+\rho^-}(m_{D^0}^2) \right],\\
\label{h5P}
h_4^P &=&  - \frac{G_F}{2}\ \Lambda_1 \   a_2\ \frac{f_{D^0}}{f_{\rho}}  A_0^{K^-\rho^+}(m_{D^0}^2),\\
\label{h6P}
h_5^P &=&  - \frac{G_F}{2}\ \Lambda_1 \   a_1.
\end{eqnarray}

The definitions and values of all quantities entering Eqs.~(\ref{h1ST}) to (\ref{h6P}) will be found in Ref.~\cite{JPD_inprogress}.


\begin{thebibliography}{99}

\bibitem{Ref1LHCb}
 R.~Aaij {\it et al.} [LHCb Collaboration], Measurements of $CP$ violation in the three-body phase space of charmless $B^{\pm}$ decays,   Phys.\ Rev.\ D {\bf 90},  112004 (2014).

\bibitem{Ref2LHCb}  LHCb Collaboration, LHCB-CONF-2011-059, Relative branching ratio measurements of charmless $B^{\pm}$ decays to three hadrons.



\bibitem{1412.4269}
J. Libby, Direct CP violation in hadronic B decays, arXiv:1412.4269v1 [hep-ex]. 

\bibitem{Ref4LHCb}
R.~Aaij {\it et al.} [LHCb Collaboration], Search for CP violation in $D^0 \to \pi^- \pi^+ \pi^0$ decays with the energy test,  Phys.\ Lett.\ B {\bf 740}, 158 (2015).



\bibitem{Ref5LHCb}
R.~Aaij {\it et al.} [LHCb Collaboration], Search for CP violation in the decay $D^+ \to \pi^-\pi^+\pi^+$,  Phys.\ Lett.\ B {\bf 728}, 585 (2014).







 \bibitem{fkll} 
A. Furman, R. Kami\'nski, L.~Le\'sniak and B.~Loiseau, Long-distance effects and final state interactions in $B \to \pi\pi K$ and $B \to K\bar  K K$ decays, Phys. Lett.  B {\bf 622}, 207 (2005).

\bibitem{ElBennich2006}
  B. El-Bennich, A. Furman, R. Kami\'nski, L.~Le\'sniak and B.~Loiseau, Interference between $f_0(980)$ and $\rho(770)^0$ resonances in $B\to\pi^+\pi^-K$ decays, Phys. Rev. D \textbf{74}, 114009 (2006).

\bibitem{ElBennichetal09}
B.~El-Bennich, A.~Furman, R.~Kami\'nski, L.~Le\'sniak, B.~Loiseau and B.~Moussallam, CP violation and kaon-pion interactions in $B \to K \pi^+ \pi^-$ decays, Phys.\ Rev.\ D {\bf 79}, 094005 (2009); Erratum-ibid, Phys.\ Rev.\ D {\bf 83}, 039903 (2011). 

\bibitem{Dedonderetal2011}
J.-P. Dedonder, A. Furman, R Kami\'nski, L. Le\'sniak and B. Loiseau, Final state interactions and CP violation in $B^{\pm}\to \pi^+ \pi^- \pi^{\pm}$ decays, Acta Phys. Pol. B {\bf 42}, 2013 (2011).

\bibitem{PLB699_102}
A. Furman, R. Kami\'nski, L. Le\'sniak, P. \.{Z}enczykowski, Final state interactions in $B^\pm \to K^+K^- K^\pm$ decays, Phys. Lett.  B {\bf 699}, 102 (2011).

\bibitem{LZ2014}
 L.~Le\'sniak and P.~\.{Z}enczykowski, Dalitz-plot dependence of CP asymmetry in $ B^{\pm} \to K^\pm K^+K^-$ decays,  Phys.\ Lett.\ B {\bf 737}, 201 (2014).

\bibitem{Boitoetal09}
D.~R.~Boito, J.-P.~Dedonder, B.~El-Bennich, O.~Leitner and B.~Loiseau, Scalar resonances in a unitary $\pi \pi$ $S$-wave model for $D^+ \to \pi^+ \pi^- \pi^+$, Phys.\ Rev.\ D {\bf 79}, 034020 (2009).
 


\bibitem{BE2009}
D.~R.~Boito and R.~Escribano, $K\pi$ form-factors and final state interactions in $D^+ \to K^- \pi^+ \pi^+$ decays, Phys. Rev. D {\bf 80}, 054007 (2009).

\bibitem{JPD_PRD89}
J.-P. Dedonder, R. R Kami\'nski, L. Le\'sniak and B. Loiseau, Dalitz plot studies of $D^0 \to K_S^0 \pi^+ \pi^-$ decays in a factorization approach, Phys. Rev. D {\bf 89}, 094018 (2014).

\bibitem{LHCbWS} 
J.~H.~Alvarenga Nogueira {\it et al.}, Summary of the 2015 LHC$b$ workshop on multi-body decays of $D$ and $B$ mesons, arXiv:1605.03889 [hep-ex].
 
 \bibitem{Beneke2003} 
M.~Beneke and M.~Neubert, QCD factorization for $B\to PP$ and $B\to PV$ decays, Nucl. Phys. B \textbf{675}, 333 (2003).


\bibitem{1308.5139}
H.~Y.~Cheng and C.~K.~Chua, Branching fractions and direct CP violation in charmless three-body decays of $B$ mesons,
  Phys.\ Rev.\ D {\bf 88}, 114014 (2013).


\bibitem{1401.5514}
 H.~Y.~Cheng and C.~K.~Chua, Charmless three-body decays of $B_s$ mesons,   Phys.\ Rev.\ D {\bf 89},  074025 (2014).

\bibitem{1502.05483} 
  W.~F.~Wang, H.~n.~Li, W.~Wang and C.~D.~L\"u, $S$-wave resonance contributions to the $B^0_{(s)}\to J/\psi\pi^+\pi^-$ and $B_s\to\pi^+\pi^-\mu^+\mu^-$ decays,  Phys.\ Rev.\ D {\bf 91},  094024 (2015).

\bibitem{1609.04614}
W.~F.~Wang and H.~n.~Li, Quasi-two-body decays $B\to K\rho\to K\pi\pi$ in perturbative QCD approach,  Phys.\ Lett.\ B {\bf 763}, 29 (2016).


\bibitem{Bauer1987} 
M. Bauer, B. Stech and M. Wirbel, Exclusive nonleptonic Decays of $D$-, $D_s$- and $B$-Mesons, Z. Phys. C {\bf 34}, 103 (1987).

\bibitem{Abbasetal} 
  A.~Biswas, N.~Sinha and G.~Abbas, Nonleptonic decays of charmed mesons into two pseudoscalars, 
  Phys.\ Rev.\ D {\bf 92},  014032 (2015). 

\bibitem{JPD_inprogress}
J.-P. Dedonder, R. R Kami\'nski, L. Le\'sniak and B. Loiseau, Dalitz plot studies of $D^0 \to K_S^0 K^+ K^-$ decays in a factorization approach,  work in progress.


\bibitem{KMV2015}
Susanne Kr\"ankl, Thomas Mannel, Javier Virto, Three-body nonleptonic $B$ decays and QCD factorization, 
Nucl. Phys. B {\bf 899}, 247 (2015). 


\bibitem{MG02}  S.~Gardner and Ulf-G.~Mei\ss ner, Rescattering and chiral dynamics in $B \to \rho \pi$ decay, Phys.\ Rev.\ D {\bf 65}, 094004 (2002).
  
\bibitem{BurasNPB434_606} A. J. Buras, QCD factors $a_1$ and $a_2$ beyond leading logarithms versus factorization in non leptonic heavy meson decays,  Nucl. Phys. B {\bf 434}, 606 (1995).

\bibitem{ElBennich:2010ha} 
  B.~El-Bennich, M.~A.~Ivanov and C.~D.~Roberts, Strong $D^* \to D \pi$ and $B^* \to B \pi$ couplings, 
  Phys.\ Rev.\ C {\bf 83}, 025205 (2011). 

\bibitem{ElBennich:2011py} 
  B.~El-Bennich, G.~Krein, L.~Chang, C.~D.~Roberts and D.~J.~Wilson, Flavor SU(4) breaking between effective couplings, Phys.\ Rev.\ D {\bf 85}, 031502 (2012). 

\bibitem{ElBennich:2012tp} 
  B.~El-Bennich, C.~D.~Roberts and M.~A.~Ivanov, Heavy-quark symmetries in the light of nonperturbative QCD approaches,  PoS QCD{\bf -TNT-II}, 018 (2012).

\bibitem{Bashir:2012fs} 
  A.~Bashir, L.~Chang, I.~C.~Clo\"et, B.~El-Bennich, Y.~X.~Liu, C.~D.~Roberts and P.~C.~Tandy, Collective perspective on advances in Dyson-Schwinger Equation QCD, 
  Commun.\ Theor.\ Phys.\  {\bf 58}, 79 (2012), 

\bibitem{Ali1998}
A. Ali, G. Kramer  and Cai-Dian L\"u, Experimental tests of factorization in charmless nonleptonic two-body $B$ decays, Phys. Rev. D \textbf{58}, 094009 (1998).

\bibitem{Beneke:2001ev}
  M.~Beneke, G.~Buchalla, M.~Neubert and C.~T.~Sachrajda, QCD factorization in $B \to \pi K, \pi \pi$ decays and extraction of Wolfenstein  parameters,  Nucl.\ Phys.\  {\bf B606}, 245 (2001).

\bibitem{Buchalla:1995vs} 
  G.~Buchalla, A.~J.~Buras and M.~E.~Lautenbacher, Weak decays beyond leading logarithms,
  Rev.\ Mod.\ Phys.\  {\bf 68}, 1125 (1996). 
 
 \bibitem{B0semilep} R.~Aaij {\it et al.} [LHCb Collaboration],
  Differential branching fraction and angular moments analysis of the decay $B^0 \to K^+ \pi^- \mu^+ \mu^-$ in the $K^*_{0,2}(1430)^0$ region,
  JHEP {\bf 1612}, 065 (2016).
  
\bibitem{D0semilep}
 R. Aaij \textit{et al.}, LHCb collaboration, First observation of the decay $D^0 \to K^- \pi^+ \mu^+ \mu^-$ in the $\rho^0$-$\omega$ region of the dimuon mass spectrum, Phys. Lett. B \textbf{757} (2016) 558. 
 
 
\bibitem{MW14} 
  U.-G.~Mei\ss ner and W.~Wang, Generalized heavy-to-light form factors in light-cone sum rules,
  Phys.\ Lett.\ B {\bf 730}, 336 (2014).

\bibitem{MW14n2}  U.~G.~Mei\ss ner and W.~Wang, $ B_s\to K^{(*)} \ell\bar \nu$, Angular Analysis, S-wave Contributions and ${ |V_{ub}|}$,
  JHEP {\bf 1401}, 107 (2014).

\bibitem{WZ15}   W.~Wang and R.~L.~Zhu, To understand the rare decay $B_s\to\pi^+\pi^-\ell^+\ell^-$,
  Phys.\ Lett.\ B {\bf 743}, 467 (2015).

\bibitem{SW15} 
Y.~J.~Shi and W.~Wang, Chiral Dynamics and $S$-wave contributions in semileptonic $D_s/B_s$ decays into $\pi^+\pi^-$,
  Phys.\ Rev.\ D {\bf 92},  074038 (2015).

\bibitem{Lietal16} 
Y.~Li, A.~J.~Ma, W.~F.~Wang and Z.~J.~Xiao, The $S$-wave resonance contributions to the three-body decays $B^0_{(s)}\rightarrow \eta _c f_0(X)\rightarrow \eta _c\pi ^+\pi ^-$ in perturbative QCD approach, 
  Eur.\ Phys.\ J.\ C {\bf 76}, 675 (2016).

\bibitem{SWZ17} 
Y.~J.~Shi, W.~Wang and S.~Zhao, Chiral dynamics, S-wave contributions and angular analysis in $D\rightarrow \pi \pi \ell \bar{\nu }$,
  Eur.\ Phys.\ J.\ C {\bf 77},  452 (2017).

 
\bibitem{Kleinetal} R.~Klein, T.~Mannel, J.~Virto and K.~K.~Vos, CP violation in multibody $B$ decays from QCD factorization,
  JHEP {\bf 1710}, 117 (2017).


 
\bibitem{Aubert:2009}
B. Aubert, \textsl{et al.} (BABAR Collaboration), Dalitz-plot analysis of $B^\pm \to \pi^\pm \pi^\mp \pi^\pm$ decays, Phys. Rev. D \textbf{79}, 072006 (2009).


\bibitem{Moussallam_2000}
B.~Moussallam, $N_f$ dependence of the quark condensate from a chiral sum rule,  Eur.\ Phys.\ J.\  C {\bf 14}, 111 (2000).

\bibitem{Fujikawa_PRD78_072006}
M. Fujikawa {\it et al.} (Belle Collaboration), High-statistics study of the $\tau^-\to\pi^-\pi^0\nu_\tau$ decay, Phys. Rev. D {\bf 78}, 072006 (2008).

\bibitem{Hanhart}
C. Hanhart, A new parametrization for the vector pion form factor, Phys. Lett. \textbf{B} 715, 170 (2012). 

\bibitem {Cheng0704.1049} 
H-Y. Cheng, C-K. Chua and A. Soni, Charmless three-body decays of $B$ mesons,  Phys. Rev D \textbf{76}, 094006 (2007). 

\bibitem{Belle}
A. Garmash \textit{et al.} (Belle Collaboration),  Dalitz analysis of the three-body charmless decays $B^+\to K^+\pi^+\pi^-$ and $B^+\to K^+K^+K^-$, Phys. Rev. D {\bf 71}, 092003 (2005).

\bibitem{BABAR}
B. Aubert\textit{ et al.} (BABAR Collaboration), Dalitz plot analysis of the decay $B^\pm \to K^\pm K^\pm K^\mp$, Phys. Rev. D {\bf 74}, 032003 (2006).

\bibitem{E791Dp3pi} E.~M.~Aitala {\it et al.} [E791 Collaboration],
   Experimental evidence for a light and broad scalar resonance in $D^+ \to \pi^- \pi^+ \pi^+$ decay,
  Phys.\ Rev.\ Lett.\  {\bf 86}, 770 (2001).

\bibitem{PDG}C.~Patrignani {\it et al.} (Particle Data Group), Review of Particle Physics,
  Chin.\ Phys.\ C {\bf 40}  100001 (2016).


\bibitem{CLEODp3pi} G.~Bonvicini {\it et al.} [CLEO Collaboration],
  Dalitz plot analysis of the $D^+ \to \pi^- \pi^+ \pi^+$ decay,''
  Phys.\ Rev.\ D {\bf 76}, 012001 (2007).


\bibitem{E791DKpipi}E.~M.~Aitala {\it et al.} [E791 Collaboration],
  Dalitz plot analysis of the decay $D^+ \to K^- \pi^+ \pi^+$ and indication of a low-mass scalar $K\pi$ resonance,
  Phys.\ Rev.\ Lett.\  {\bf 89}, 121801 (2002).

\bibitem{FOCUSDKpipi} J.~M.~Link {\it et al.} [FOCUS Collaboration],
  Dalitz plot analysis of the $D^{+} \to K^{-} \pi^{+} \pi^{+}$ decay in the FOCUS experiment,  Phys.\ Lett.\ B {\bf 653}, 1 (2007).

\bibitem{CLEODKpipi}G.~Bonvicini {\it et al.} [CLEO Collaboration],
  Dalitz plot analysis of the $D^+ \to K^- \pi^+ \pi^+$ decay,
  Phys.\ Rev.\ D {\bf 78}, 052001 (2008).


\bibitem{Magetal2011}
P. C. Magalh\~aes, M. R. Robilotta, K. S. F. F. Guimar\~aes, T. Frederico, W. de Paula, I. Bediaga, A. C. dos Reis, C. M. Maekawa, G. R. S. Zarnauskas, Towards three-body unitarity in $D^+ \to K^- \pi^+ \pi^+$, 
   Phys.\ Rev.\ D {\bf 84}, 094001 (2011).

\bibitem{MR15}
P.~C.~Magalh\~aes and M.~R.~Robilotta, $D^+ \to K^- \pi^+ \pi^+$ - the weak vector current,  Phys.\ Rev.\ D {\bf 92},  094005 (2015).


\bibitem{NKS12} 
 F.~Niecknig, B.~Kubis and S.~P.~Schneider,
  Dispersive analysis of $\omega \to 3\pi$ and $\phi \to 3\pi$ decays,
  Eur.\ Phys.\ J.\ C {\bf 72}, 2014 (2012).


\bibitem{NK15} 
F.~Niecknig and B.~Kubis,  Dispersion-theoretical analysis of the $D^{+}\to K^-\pi^{+} \pi^{+}$ Dalitz plot,
  JHEP {\bf 1510}, 142 (2015).

\bibitem{NK17} 
F.~Niecknig and B.~Kubis,
 Consistent Dalitz plot analysis of Cabibbo-favored $D^+ \to \bar{K} \pi \pi^+$ decays,  arXiv:1708.00446 [hep-ph].


\bibitem{Nakamura} 
S.~X.~Nakamura, Coupled-channel analysis of $D^+\to K^- \pi^+\pi^+$ decay,
  Phys.\ Rev.\ D {\bf 93},  014005 (2016).



\bibitem{A.Poluektov_PRD81_112002_Belle}
A. Poluektov \textit{et al.} (Belle Collaboration), Evidence for direct CP violation in the decay $B^\pm \to D^{(*)} K^\pm, D\to K_S^0 \pi^+ \pi^-$ and measurement of the CKM phase $\phi_3$, 
Phys. Rev. D \textbf{81}, 112002 (2010).


\bibitem{SanchezPRL105_081803}
 P. del Amo Sanchez \textit{et al.} (BABAR Collaboration), Measurement of $D^0-\bar  D^0$ Mixing Parameters Using $D^0 \to K_S^0 \pi^+ \pi^-$ and $D^0 \to K_S^0 K^+K^-$ Decays, 
 Phys. Rev. Lett. \textbf{105}, 081803 (2010). 
 
\bibitem{BLoiseau_Rio2015}
B.~Loiseau, J.-P.~Dedonder, R. R Kami\'nski and L. Le\'sniak, ``{\em Constrained amplitude analysis of some three-body hadronic $D$-decay data\/}", talk given by 
B.~Loiseau at the LHC$b$ workshop  on multi-body decays of $B$ and $D$ mesons, July 27 - 30, 2015, Rio De Janeiro, Brazil; https://indico.cern.ch/event/359085/timetable/\#all.

\bibitem{Barton65}
G.~Barton , Introduction to dispersion techniques in field theory, W. A. Benjamin, INC., New York (1965).

\bibitem{MO}
N. I. Muskhelishvili, Singular integral equations, (P.Nordhof 1953), chapters 18 and 19;
R. Omn\`es, On the Solution of certain singular integral equations of quantum field theory, 
Nuovo Cim. {\bf 8}, 316 (1958).

\bibitem{DHK16}
J. T. Daub, C. Hanhart, B. Kubis, A model-independent analysis of final-state interactions in $\bar B_{d/s}^0 \to J/\psi \pi \pi$, JHEP {\bf 1602}, 009 (2016). 

\bibitem {meis01}
Ulf-G.~Meissner and J.~A.~Oller, $J/\Psi \to \phi \pi \pi (K\bar K)$ decays, chiral dynamics and OZI violation, Nucl. Phys. \textbf{A679}, 671 (2001).
 
\bibitem{BelleDKspipi} K.~Abe {\it et al.} [Belle Collaboration],
  Measurement of $D_0 - \bar D_0$ Mixing Parameters in $D_0 \to K_{(s)}^0 \pi^+ \pi^-$ decays,
  Phys.\ Rev.\ Lett.\  {\bf 99}, 131803 (2007).


 \bibitem{Kaminski_EPJC9_141}
  R.~Kami\'nski, L.~Le\'sniak and B.~Loiseau, Scalar mesons and multichannel amplitudes, 
 Eur. Phys. J. {\bf C9}, 141 (1999).

\bibitem{RoigDumm}D.~G\'omez Dumm and P.~Roig,
  Dispersive representation of the pion vector form factor in $\tau\to\pi\pi\nu_\tau$ decays,
  Eur.\ Phys.\ J.\ C {\bf 73}, 2528 (2013).

\bibitem{Moussallam:2007qc}B.~Moussallam, Analyticity constraints on the strangeness changing vector current and applications to $\tau \to K \pi \nu_\tau$ and $\tau \to K \pi \pi \nu_\tau$, 
    Eur.\ Phys.\ J.\ C {\bf 53}, 401 (2008). 
   
 \bibitem{JOP}   M.~Jamin, J.~A.~Oller and A.~Pich, Light quark masses from scalar sum rules, Eur.\ Phys.\ J.\ C {\bf 24}, 237 (2002);
  M.~Jamin, J.~A.~Oller and A.~Pich, Scalar $K \pi$ form factor and light quark masses, Phys.\ Rev.\ D {\bf 74}, 074009 (2006).

\bibitem{SDGBM06}
S. Descotes-Genon, B. Moussallam, The $K^*_0(800)$ scalar resonance from Roy-Steiner representations of $\pi K$ scattering, Eur. Phys. J. C {\bf48}, 553 (2006).

\bibitem{JOPKpiScattering}
M.~Jamin, J.~A.~Oller and A.~Pich,
  $S$-wave $K\pi$ scattering in chiral perturbation theory with resonances,
  Nucl.\ Phys.\ B {\bf 587}, 331 (2000).

  
\bibitem{Boito:2008fq} 
  D.~R.~Boito, R.~Escribano and M.~Jamin,
  $K \pi$ vector form-factor, dispersive constraints and $\tau \to  K\pi\nu_\tau$ decays,
  Eur.\ Phys.\ J.\ C {\bf 59}, 821 (2009).
    
\bibitem{Boito:2010me} 
  D.~R.~Boito, R.~Escribano and M.~Jamin,
  $K\pi$ vector form factor constrained by $\tau \to  K\pi \nu_\tau$ and $K_{l3}$ decays,
  JHEP {\bf 1009}, 031 (2010).

\bibitem{BelleTauKpinu}   D.~Epifanov {\it et al.} [Belle Collaboration],
  Study of $\tau^- \to K_S \pi^- \nu_\tau$ decay at Belle,
  Phys.\ Lett.\ B {\bf 654}, 65 (2007).

\bibitem{EPJC_75_488}
M.~Albaladejo, B.~Moussallam, Form factors of the isovector scalar current and the $\eta \pi$ scattering phase shifts, Eur.\ Phys.\ J.\ C {\bf 75}, 488 (2015).

\bibitem{Bruch}
C. Bruch, A. Khodjamirian, J. H. K\"uhn, Modeling the pion and kaon form factors in the timelike region, Eur. Phys. J. {\bf C39}, 41 (2005). 

\bibitem{Lu:2007sg} 
  C.~D.~Lu, W.~Wang and Z.~T.~Wei,
  Heavy-to-light form factors on the light cone,
  Phys.\ Rev.\ D {\bf 76}, 014013 (2007).
  
\bibitem{Melikhov:2001zv} 
  D.~Melikhov,
  Dispersion approach to quark binding effects in weak decays of heavy mesons,
  Eur.\ Phys.\ J.\ direct {\bf 4},  2 (2002).
  
\bibitem{Ivanov:2007cw} 
  M.~A.~Ivanov, J.~G.~K\"orner, S.~G.~Kovalenko and C.~D.~Roberts,
  $B$-meson to light meson transition form-factors,
  Phys.\ Rev.\ D {\bf 76}, 034018 (2007).
 
 \bibitem{Khodjamirian:2006st} 
  A.~Khodjamirian, T.~Mannel and N.~Offen,
  Form-factors from light-cone sum rules with $B$-meson distribution amplitudes,
  Phys.\ Rev.\ D {\bf 75}, 054013 (2007).
  
 \bibitem{Ball:2004rg} 
  P.~Ball and R.~Zwicky,
  $B_{d,s} \to  \rho, \omega, K^*, \phi$ decay form-factors from light-cone sum rules revisited,
  Phys.\ Rev.\ D {\bf 71}, 014029 (2005).
  
 \bibitem{Leitner:2010fq} 
  O.~Leitner, J.-P.~Dedonder, B.~Loiseau and B.~El-Bennich,
  Scalar resonance effects on the $B_{s} - \bar B_{s}$ mixing angle,
  Phys.\ Rev.\ D {\bf 82}, 076006 (2010).
  
  \bibitem{Paracha:2014wra} 
  M.~A.~Paracha, B.~El-Bennich, M.~J.~Aslam and I.~Ahmed,
  Ward identities, ${ B\to V}$ transition form factors and applications,
  J.\ Phys.\ Conf.\ Ser.\  {\bf 630},  012050 (2015).
  
  
  \bibitem{Lees:2014ihu} 
  J.~P.~Lees {\it et al.} [BaBar Collaboration],
  Measurement of the $D^0 \to \pi^- e^+ \nu_e$ differential decay branching fraction as a function of $q^2$ and study of form factor parameterizations,
  Phys.\ Rev.\ D {\bf 91}, 052022 (2015).
  
  \bibitem{ElBennich:2009vx}
  B.~El-Bennich, M.~A.~Ivanov and C.~D.~Roberts,
  Flavourful hadronic physics,
  Nucl.\ Phys.\ Proc.\ Suppl.\  {\bf 199}, 184  (2010).






\end{thebibliography}
\end{document}